\documentclass[a4paper]{elsart5p}
\usepackage{txfonts}
\usepackage{natbib,graphicx,url,rotating}
\usepackage{hyperref}
\usepackage{amssymb}
\begin{document}
%
%
\newcommand{\refcomment}[1]{[ #1 ]}
\newcommand{\OB}{\ensuremath{\Omega_{\mathrm{A}}}}
\newcommand{\Bnu}{\ensuremath{B_{\nu}}}
\newcommand{\Snu}{\ensuremath{S_{\nu}}}
\newcommand{\Tb}{\ensuremath{T_{\mathrm{B}}}}
\newcommand{\Tsys}{\ensuremath{T_{\mathrm{sys}}}}
\newcommand{\Tsky}{\ensuremath{T_{\mathrm{sky}}}}
\newcommand{\TRMS}{\ensuremath{T_{\mathrm{RMS}}}}
\newcommand{\texp}{\ensuremath{t_{\mathrm{exp}}}}
\newcommand{\Aeff}{\ensuremath{A_{\mathrm{eff}}}}
\newcommand{\Dmax}{\ensuremath{D_{\mathrm{max}}}}
\newcommand{\lmax}{\ensuremath{l_{\mathrm{max}}}}
\newcommand{\de}{\ensuremath{\mathrm{d}}}
\newcommand{\gff}{\ensuremath{g_{\mathrm{ff}}}}
\newcommand{\arcmin}{\ensuremath{^\prime}}
\newcommand{\arcsec}{\ensuremath{^{\prime\prime}}}
\newcommand{\la}{\lesssim}
\newcommand{\ga}{\gtrsim}
\renewcommand{\deg}{\ensuremath{^\circ}}
\newcommand{\farcs}{\ensuremath{.\!\!^{\prime\prime}}}
\newcommand{\fdg}{\ensuremath{.\!\!^\circ}}

\newcommand{\urlfn}[1]{\footnote{\url{#1}}}
\makeatletter
\newcommand\ion[2]{#1\,{\small\rmfamily\@Roman{#2}}}
%
%
%
%


\let\jnl@style=\rm
\def\ref@jnl#1{{\jnl@style#1 }}

\def\aj{\ref@jnl{AJ}}                   
\def\araa{\ref@jnl{ARA\&A}}             
\def\apj{\ref@jnl{ApJ}}                 
\def\apjl{\ref@jnl{ApJ}}                
\def\apjs{\ref@jnl{ApJS}}               
\def\ao{\ref@jnl{Appl.~Opt.}}           
\def\apss{\ref@jnl{Ap\&SS}}             
\def\aap{\ref@jnl{A\&A}}                
\def\aapr{\ref@jnl{A\&A~Rev.}}          
\def\aaps{\ref@jnl{A\&AS}}              
\def\azh{\ref@jnl{AZh}}                 
\def\baas{\ref@jnl{BAAS}}               
\def\jrasc{\ref@jnl{JRASC}}             
\def\memras{\ref@jnl{MmRAS}}            
\def\mnras{\ref@jnl{MNRAS}}             
\def\pra{\ref@jnl{Phys.~Rev.~A}}        
\def\prb{\ref@jnl{Phys.~Rev.~B}}        
\def\prc{\ref@jnl{Phys.~Rev.~C}}        
\def\prd{\ref@jnl{Phys.~Rev.~D}}        
\def\pre{\ref@jnl{Phys.~Rev.~E}}        
\def\prl{\ref@jnl{Phys.~Rev.~Lett.}}    
\def\pasp{\ref@jnl{PASP}}               
\def\pasj{\ref@jnl{PASJ}}               
\def\qjras{\ref@jnl{QJRAS}}             
\def\skytel{\ref@jnl{S\&T}}             
\def\solphys{\ref@jnl{Sol.~Phys.}}      
\def\sovast{\ref@jnl{Soviet~Ast.}}      
\def\ssr{\ref@jnl{Space~Sci.~Rev.}}     
\def\zap{\ref@jnl{ZAp}}                 
\def\nat{\ref@jnl{Nature}}              
\def\iaucirc{\ref@jnl{IAU~Circ.}}       
\def\aplett{\ref@jnl{Astrophys.~Lett.}} 
\def\apspr{\ref@jnl{Astrophys.~Space~Phys.~Res.}}
\def\bain{\ref@jnl{Bull.~Astron.~Inst.~Netherlands}} 
\def\fcp{\ref@jnl{Fund.~Cosmic~Phys.}}  
\def\gca{\ref@jnl{Geochim.~Cosmochim.~Acta}}   
\def\grl{\ref@jnl{Geophys.~Res.~Lett.}} 
\def\jcp{\ref@jnl{J.~Chem.~Phys.}}      
\def\jgr{\ref@jnl{J.~Geophys.~Res.}}    
\def\jqsrt{\ref@jnl{J.~Quant.~Spec.~Radiat.~Transf.}}
\def\memsai{\ref@jnl{Mem.~Soc.~Astron.~Italiana}}
\def\nphysa{\ref@jnl{Nucl.~Phys.~A}}   
\def\physrep{\ref@jnl{Phys.~Rep.}}   
\def\physrev{\ref@jnl{Phys.~Rev.}}   
\def\physscr{\ref@jnl{Phys.~Scr}}   
\def\planss{\ref@jnl{Planet.~Space~Sci.}}   
\def\procspie{\ref@jnl{Proc.~SPIE}}   
\def\rmp{\ref@jnl{Rev.~Mod.~Phys.}} 	

\let\astap=\aap
\let\apjlett=\apjl
\let\apjsupp=\apjs
\let\applopt=\ao

\makeatother
\begin{frontmatter}
\title{Science with a lunar low-frequency array:\\ 
From the dark ages of the Universe to nearby exoplanets}
\author{Sebastian Jester}
\ead{jester@mpia.de}
\address{Max-Planck-Institut f\"ur Astronomie, K\"onigstuhl 17,
  69117 Heidelberg, Germany}
\author{Heino Falcke}
\ead{H.Falcke@astro.ru.nl}
\address{Department of Astrophysics, Institute
  for Mathematics, Astrophysics and Particle Physics, Radboud
  University, P.O. Box 9010, 6500 GL Nijmegen, The Netherlands}
\address{ASTRON, Oude Hoogeveensedijk 4, 7991 PD Dwingeloo, The Netherlands}

\begin{abstract}
  Low-frequency radio astronomy is limited by severe ionospheric
  distortions below 50\,MHz and complete reflection of radio waves
  below 10--30 MHz. Shielding of man-made interference from long-range
  radio broadcasts, strong natural radio emission from the Earth's
  aurora, and the need for setting up a large distributed antenna
  array make the lunar far side a supreme location for a low-frequency
  radio array. A number of new scientific drivers for such an array,
  such as the study of the dark ages and epoch of reionization,
  exoplanets, and ultra-high energy cosmic rays, have emerged and need
  to be studied in greater detail.  Here we review the scientific
  potential and requirements of these new scientific drivers and
  discuss the constraints for various lunar surface arrays.  In
  particular we describe observability constraints imposed by the
  interstellar and interplanetary medium, calculate the achievable
  resolution, sensitivity, and confusion limit of a dipole array using
  general scaling laws, and apply them to various scientific
  questions.
  
  Of particular interest for a lunar array are studies of the earliest
  phase of the universe which are not easily accessible by other
  means. These are the \emph{epoch of reionization} at redshifts
  z=6-20, during which the very first stars and galaxies ionized most
  of the originally neutral intergalactic hydrogen, and the \emph{dark
    ages} prior to that.

  For example, a global 21-cm wave absorption signature from
  primordial hydrogen in the dark ages at $z=30$--50 could in
  principle be detected by a single dipole in an eternally dark crater
  on the moon, but foreground subtraction would be extremely
  difficult. Obtaining a high-quality power spectrum of density
  fluctuations in the epoch of reionization at $z=6$--30, providing a
  wealth of cosmological data, would require about $10^3$--$10^5$
  antenna elements on the moon, which appears not unreasonable in the
  long term. Moreover, baryonic acoustic oscillations in the dark ages
  at z=30-50 could similarly be detected, thereby providing pristine
  cosmological information, e.g., on the inflationary phase of the
  universe.

  With a large array also exoplanet magnetospheres could be detected
  through Jupiter-like coherent bursts. Smaller arrays of order $10^2$
  antennas over $\sim$ 100\,km, which could already be erected
  robotically by a single mission with current technology and
  launchers, could tackle surveys of steep-spectrum large-scale radio
  structures from galaxy clusters and radio galaxies. Also, at very
  low frequencies the structure of the interstellar medium can be
  studied tomographically. Moreover, radio emission from neutrino
  interactions within the moon can potentially be used to create a
  neutrino detector with a volume of several cubic kilometers. An
  ultra-high energy cosmic ray detector with thousands of square
  kilometre area for cosmic ray energies $>10^{20}$\,eV could in
  principle be realized with some hundred antennas.

  In any case, pathfinder arrays are needed to test the feasibility of
  these experiments in the not too distant future.  Lunar
  low-frequency arrays are thus a timely option to consider, offering
  the potential for significant new insights into a wide range of
  today's crucial scientific topics. This would open up one of the
  last unexplored frequency domains in the electromagnetic spectrum.
\end{abstract}

\begin{keyword}
Instrumentation: interferometers \sep Cosmological parameters \sep
Neutrinos \sep Surveys

\PACS 95.55.Jz \sep  95.80.+p \sep 95.85.Ry  \sep 98.80.-k 
\end{keyword}
\end{frontmatter}

\section{Introduction}
\label{s:intro}
Low-frequency radio astronomy is currently experiencing an impressive
revival with a number major new facilities in operation or under
construction, such as the Giant Metrewave Radio Telescope
\citep[GMRT,][]{SKVea91}\urlfn{http://www.gmrt.ncra.tifr.res.in},
Low-Frequency ARray
\citep[LOFAR,][]{FalckeLOFAR06,RoettgeringLOFAR06}\urlfn{http://www.lofar.org},
Long-Wavelength Array
\citep[LWA,][]{KPCea05}\urlfn{http://lwa.unm.edu}, Murchison Widefield
Array \citep[MWA, formerly known as the Mileura Widefield
  Array,][]{Morales05,MBCea06}\urlfn{http://www.haystack.mit.edu/ast/arrays/mwa/},
the 21 Centimeter Array \citep[21CMA\urlfn{http://21cma.bao.ac.cn/}; formerly called \emph{Primeval
    Structure Telescope},
  PAST\urlfn{http://web.phys.cmu.edu/~past/overview.html},][]{PetersonPenWu05},
and Precision Array to Probe the Epoch of Reionization
\citep[PAPER,][]{BBPea05}. Eventually the Square-Kilometre Array
\citep[SKA,][]{Schilizzi04,CarilliRawlings04}\urlfn{http://www.skatelescope.org}
may extend the expected collecting area even further. Main science
drivers of these telescopes are the so-called epoch of ionization
\citep[e.g.,][and references therein]{ShaverdeBruyn00,CFBea04}, large
scale surveys, transient and variable source monitoring
\citep{FenderLOFAR06}, observations of our own sun \citep{Kasper06}
and solar system \citep{Zarka00}, as well as exoplanets \citep{Zarka07}
and astroparticle physics \citep{FG03}.  This rapid development
together with the lively discussion of a possible return to the lunar
surface by various space agencies, has raised again the long-standing
question about the potential of low-frequency astronomy from space
\citep[e.g., ][]{LMBea07}.

The currently planned ground-based telescopes will provide a serious
advance in radio astronomy and extend the accessible frequencies to
the widest range possible from the ground. The Earth's turbulent
ionosphere gives rise to ``radio seeing'', making ground-based radio
observations of the sky become difficult at $\nu\la 100$\,MHz. At even
longer wavelengths below about 10-30 MHz one encounters the
ionospheric cut-off where radio waves are reflected, permitting
long-distance short wave transmission around the Earth, but
prohibiting observations of the sky. Observing just above the cut-off,
i.e., between $\sim$10--50 MHz requires especially favourable
geomagnetic and ionospheric conditions to obtain any decent images.
The range below the cutoff is only readily observable from
space. Hence, the dominant ``low-frequency/long-wavelength'' regime
for which ground-based telescopes are being designed is at frequencies
$>30$\,MHz and wavelengths $<10$\,m. We will therefore hereafter refer
to wavelengths above 10\,m as the ultra-long wavelength (ULW)
range. This wavelength-based definition is preferable to a
frequency-based one to avoid confusion with the technical very low
frequency (VLF) designation that extends only from 3-30 kHz; the range
300 kHz-30 MHz is in engineering language named the medium and high
frequency regime (MF \& HF), which is clearly not appropriate in a
modern astronomical context.

The best resolution achieved so far in the ULW range is on the scale
of around 5 degrees, but more typically of order tens of degrees. This
compares rather unfavourably to the milli-arcsecond resolution that
can be routinely obtained in very long baseline interferometry (VLBI)
at higher radio frequencies. Hence, the low-frequency Universe is the
worst-charted part of the radio spectrum, and perhaps even of the
entire electromagnetic spectrum.  By today, only two kinds of maps of
the sky have been made at frequencies below 30 MHz. The first are maps
of a part of the southern sky near the Galactic center such as those
obtained by \citet{CaneWhitham77}, \citet{EllisMendillo87} and
\citet{CaneErickson01} from Tasmania. These have angular resolutions
ranging from 5 to 30 degrees. The second kind are the maps obtained by
the RAE-2 satellite \citep{NB78} with angular resolution of 30 degrees
or worse. None of these maps show individual sources other than
diffusive synchrotron emission of the Galaxy, nor do they cover the
entire sky.

To improve this situation and to overcome these limitations,
space-based low-frequency telescopes are required for all observations
below the ionospheric cutoff \citep{Weiler87,WJSea88,KassimWeiler90}.
This is also true for a significant part of the seeing-affected
frequency range above the cutoff frequency where high-resolution and
high-dynamic range observations are required, such as imaging of 21-cm
emission of neutral hydrogen in the very early Universe \citep{CHL07}.

\begin{figure*}
\begin{center}
\includegraphics[angle=270,width=0.66\textwidth]{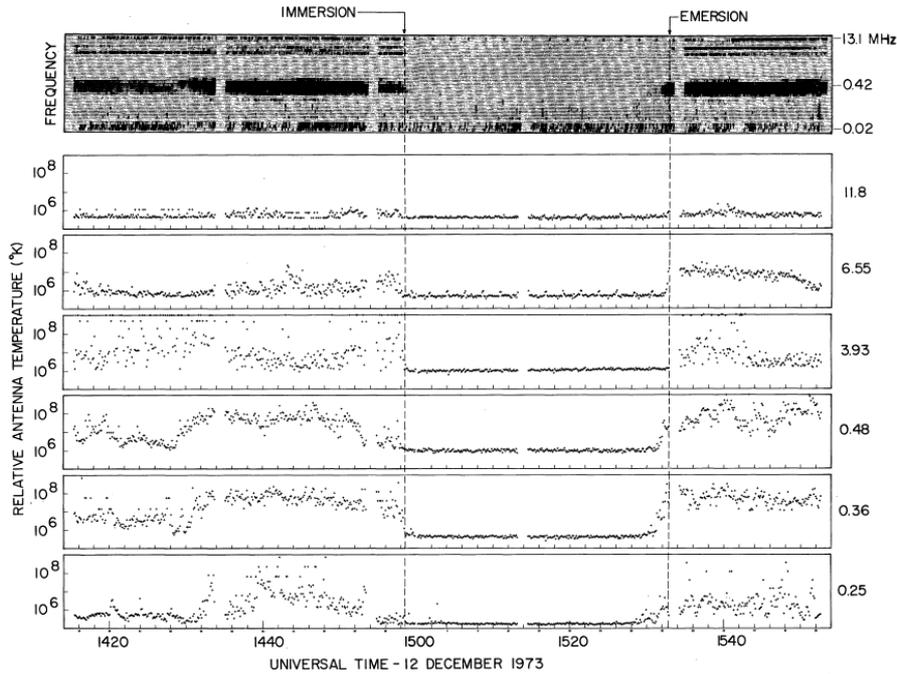}
\end{center}
\caption{\label{f:RAE2shield}Shielding of terrestrial radio
  interference by the moon, as observed by the RAE-2 satellite. Even
  at a distance of 300,000 km from the Earth, the additional shielding
  by the Moon provides 1-3 orders of magnitude (10-30 dB) of
  additional interference suppression. Taken from
  \protect\citet[\copyright\ ESO]{AKNea75}.}
\end{figure*}
So far, there have only been two space missions whose primary purpose
was low-frequency radio astronomy: the Radio Astronomy Explorers (RAE)
1 and 2. The design and results of these missions have been reviewed
by \citet{Kaiser87}. \citet{Kaiser90} summarized the science goals of
the RAE missions as four-fold, namely characterizing the spatial and
spectral structure of Galactic radio emission, observing bursts from
the Sun and Jupiter, and attempting to detect discrete Galactic and
extragalactic sources. It was a surprising finding of RAE-1 that the
Earth itself is a strong emitter of low-frequency bursts, the
so-called Auroral Kilometric Radiation (AKR) generated by solar-wind
interactions with the Earth's magnetic field. This emission is so
strong that RAE-2 was placed in a lunar instead of the originally
planned terrestrial orbit to provide shielding from this natural
terrestrial interference (Fig.~\ref{f:RAE2shield} and
\S\,\ref{s:constr.int}). Of the original science goals, only the solar
and Jupiter burst studies could be completed; in addition, terrestrial
bursts were studied. The structure of the Galaxy's emission was only
seen at very low spatial resolution (with beam sizes between
$37^{\circ}\times61^{\circ}$ and $220^{\circ}\times160^{\circ}$) and
fairly low signal-to-noise ratio. Due to the very limited angular
resolution and the large power of AKR, it proved impossible to image
any discrete sources directly. Thus, two of the initial scientific
questions of the RAE missions remain unanswered to date.

Low-frequency radio imaging telescopes at these frequencies are
nowadays realized as digitally phased arrays of thousands of simple
dipole antennas that are spread over a large area. This puts a serious
constraint on any realization consisting of free-flying antennas in
space and very likely favours lunar surface-based concepts in the long
term. Moreover the moon provides unique shielding against disturbing
radio interference from sun and Earth and within the exploration
efforts of space agencies lunar missions appear to be a realistic
possibility. Hence we will focus here on a lunar surface telescope
concept consisting of dipoles as the most basic antenna unit.
Nonetheless, most of our conclusions are general enough that they can
be of use also for other kinds of implementations.

Over the past four decades several conferences and study groups formed
by space agencies have produced reports on lunar radio astronomy
\citep{Weiler87,WJSea88,KassimWeiler90,alfis,LCDea91,LCCea92,Dainty92,Foing94,Foing96,EsaMoon97,Woan97,JWAea98,JABea00,SWGB00,Takahashi03,OP05,KPJH05}.
Starting from these earlier studies we will here summarize the
prospects and fundamental limitations of ULW observations and discuss
their scientific potential in light of our current knowledge. Recent
scientific advances in many areas and in particular in cosmology lets
low-frequency astronomy shine in a new light and makes this a timely
topic.

The plan of the paper is as follows. We describe observability
constraints for ULW observations from the Earth and the Moon in
\S\ref{s:constr}, including a description of the Galactic foreground
emission, and formulae for the estimation of array sizes,
sensitivities etc. We give science questions to be addressed by a
lunar low-frequency array in \S\ref{s:scidrv} and summarize our
findings in \S\ref{s:summary}.

\section{Observability constraints for ULW observations}
\label{s:constr}

\subsection{Earth Ionosphere}
\label{s:constr.iono}

Radio waves are scattered by free electrons, such as those occurring
in a space plasma or the Earth's ionosphere. The scattering strength
increases towards lower frequencies. At frequencies below the plasma
frequency, which increases with electron density, radio waves cannot
propagate at all. The plasma frequency of the Earth's ionosphere is
typically near 10~MHz by day and near 5~MHz by night, implying that
radio observations near and below these frequencies are impossible
from the Earth (the plasma frequencies can reach somewhat lower values
near the Earth's magnetic poles). At frequencies up to a few 100 MHz,
the ionosphere causes celestial radio sources to suffer significantly
from angular displacements, angular broadening, and intensity
fluctuations (scintillation). These features are akin to
\emph{seeing}, the twinkling of stars visible to the eye, which is
caused by density fluctuations of turbulence layers in the Earth's
atmosphere. Because of the scattering, the best available ground-based
maps of the low-frequency sky have resolutions ranging from 30 degrees
at 2.1~MHz to 5 degrees at 10~MHz
\citep{EllisMendillo87,CaneErickson01} and very poor dynamic range.

\subsection{Lunar Surface and Ionosphere}
\label{s:constr.lunar}

Even though the moon does not have a gaseous atmosphere, instruments
carried by the Apollo~14 and Luna~19 and 22 missions showed that the
moon's surface has a photoelectron sheath, i.e., a weak ionosphere. On
the day side, the inferred electron densities lay in the range
500--10,000$/$cm$^3$, with plasma frequencies between 0.2 and 1 MHz
\citetext{see \citealt{RB72}, \citealt{BFHea75} and references
  therein}. Electron densities on the night side are expected to be
much lower, since the moon's surface potential becomes
negative. However, the electron density has not been mapped yet in
detail as function of lunar position and altitude above the
surface. This is clearly a necessity, given that electron densities at
the high end of the inferred ranges would mean that a lunar
observatory only gains a factor of 10 in frequency during lunar day
compared to what is observable below the Earth's ionosphere. The
current generation of moon missions, in particular the Japanese
KAGUYA/SELENE mission, should clarify this issue.

Secondly, low-frequency radio waves can propagate into the lunar
regolith to depths of 1-100 wavelengths. Subsurface discontinuities in
the electrical properties of the lunar regolith could therefore lead
to reflections and hence stray signals being scattered into the beam
of lunar antennas. Thus, the subsurface electrical properties of the
prospective sites for a lunar radio telescope need to be explored
before construction of a very large array
\citep[compare][]{Takahashi03}.

\subsection{Man-made and natural Interference}
\label{s:constr.int}

The low-frequency radio spectrum is occupied to a large fraction by
terrestrial broadcasts, in particular FM and longer-wave radio and
television broadcasts and military and civil communications. This
radio frequency interference (RFI) is a severe problem for all kinds of
ground-based radio telescopes. These signals block out cosmic signals
directly at the corresponding wavelengths, but they also lead to an
increased noise level for observations at other frequencies through
intermodulation products in the receiving system, through leakage, or
if the affected frequencies are not filtered out completely.

At the longest wavelengths, these signals are reflected by the
ionosphere. This is of course the very reason why these wavelengths
are used for world-wide radio broadcasts, but it also means that a
terrestrial ULW radio telescope is sensitive not just to local
interference, but also to interference from all parts of the world,
irrespective of location.

When the first dedicated radio astronomy satellite, RAE-1, was
designed, it was expected that placing the satellite in a low orbit
above the ionosphere would lead to a sufficient suppression of
interference. However, it turned out that there were very strong radio
bursts from the Earth itself, generated by the interaction between
solar-wind particles and the Earth's magnetic field, in particular in
the frequency range 150-300 kHz. At higher frequencies, man-made and
lightning interference are also still detectable above the
ionosphere. Therefore, the second mission, RAE-2, was redesigned and
this satellite was put into an orbit around the Moon. Figure
\ref{f:RAE2shield} shows how well terrestrial radio interference is
shielded by the Moon.

Naturally, also the sun and Jupiter are strong sources of ULW radio
emission - especially during outbursts - and again here the moon can
act as a perfect shield for a large fraction of the time.

\begin{figure}
\begin{center}
\includegraphics[width=0.5\textwidth]{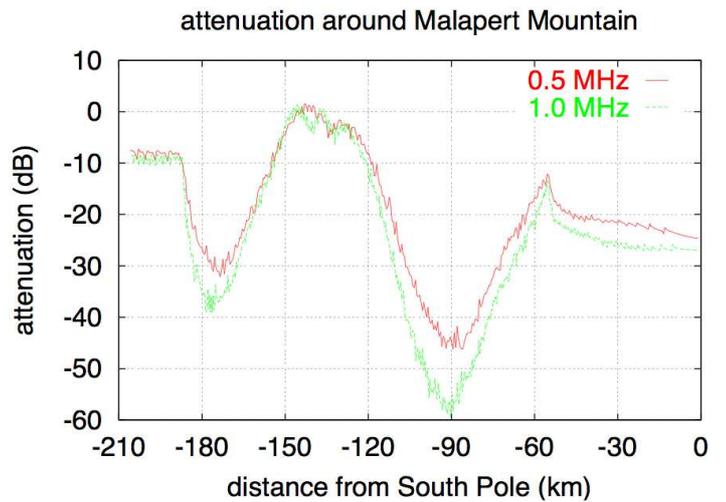}
\end{center}
\caption{\label{f:takahashi_Malapert}Shielding of terrestrial radio
  interference by Malapert mountain, whose peak is at $\approx
  120$\,km from the south pole.  This shielding is in addition to the
  20--30\,dB attenuation applying to all interference in the south
  polar region. Taken from \protect\citet[Fig.~4.9]{Takahashi03} with
  kind permission of the author.}
\end{figure}
\citet{Takahashi03} has modelled the reflection and absorption of
low-frequency radio waves by the Moon, including the shielding by
mountains.  He finds that a large fraction of the incident energy is
reflected by the lunar surface (55\% at 0.5\,MHz), and an effective
skin depth of $2.2\,\mathrm{km}\, (\nu/\mathrm{MHz})^{-0.81}$.  Thus, at
0.5\,MHz, shielding is already substantial at a depth of a few
kilometers and becomes better at higher frequencies.
Figure~\ref{f:takahashi_Malapert} shows the attenuation along a line
joining the lunar south pole (a possible location for a future lunar
base) and the nearby Malapert mountain; Malapert may provide up to
70\,dB (a factor of $10^7$) of shielding at 0.5\,MHz, while the
simulations indicate a shielding by a factor of at least $10^8$ at all
frequencies above 10\,kHz for the lunar far side, i.e., substantially
stronger shielding.

\subsection{The interplanetary and interstellar media}
\label{s:constr.media}

A fundamental constraint for all very low-frequency radio astronomy is
the presence of free electrons in the interplanetary medium (IPM)
within the solar system, the solar wind, and in the interstellar
medium (ISM) filling our Galaxy. The plasma frequency of the IPM
depends on the distance from the sun and the solar cycle; at the
Earth, it lies in the range 20-30~kHz, and the scaling is roughly
linear with sun-Earth distance.  Observations outside the solar system
are restricted to higher frequencies. The ISM plasma frequency is of
order 2~kHz for the average electron density in the Galactic plane of
0.025\,cm$^{-3}$ \citep{PW02} and therefore poses no additional
constraints beyond those imposed by the IPM. The ISM plasma frequency
is significantly lower than the ionospheric plasma frequency and hence
there is still a wide range of accessible parameter space to be
explored.

However, in addition to inhibiting the propagation of electromagnetic
waves below the plasma frequency, there are several more consequences
of the presence of free electrons: foreground emission, free-free
absorption, and scattering of waves in turbulent regions. As usual all
these effects can be a curse, when studying extragalactic sources, and
a blessing, when studying the local ISM. Moreover, there are
fundamental constraints to observations of the radio sky with
interferometers, which arise from the need to match the resolution to
the density of sources at the instrument's sensitivity to avoid
confusion. We consider each of these effects in turn.

\subsubsection{Galactic synchrotron emission}
\label{s:constr.gal.syn}

\begin{figure}
\begin{center}
\includegraphics[angle=270,width=0.5\textwidth]{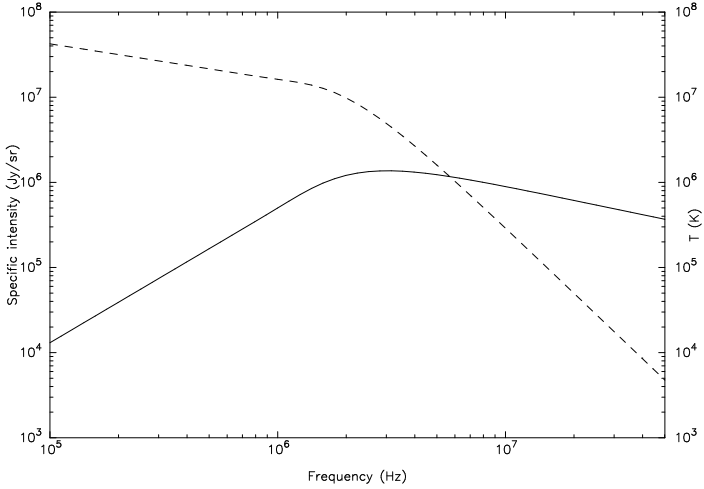}
\end{center}
\caption{\label{f:galBG}Spectrum of Galactic synchrotron emission,
  showing specific intensity (solid line) and surface brightness
  temperature (dashed line). The surface brightness temperature
  dominates the system temperature at the frequencies shown. Figure
  taken from \protect\citet{OP03} with kind permission of the authors.}
\end{figure}
Synchrotron emission from electrons moving in the Galactic magnetic
field constitutes the dominant foreground (an overview about
observations of the Galactic emission that have been made to date is
given by \citealt{Dwarakanath00} and \citealt{KWL04}). The
distributions of electrons and magnetic fields are not yet known in
full detail and subject of active study. The component of the ISM that
contains the free electrons responsible for the scattering is called
the Warm Ionized Medium (WIM). It broadly follows the distribution of
gas in the Galaxy, i.e., it is concentrated in the plane of the
Galaxy. The brightness temperature of the emission rises from about
$10^{4}$\,K at 30~MHz to about $2.6\times 10^{7}\,$K at around 1~MHz,
and then increases less rapidly (and the specific intensity decreases)
because free electrons absorb low-frequency radiation via free-free
absorption (see Figure~\ref{f:galBG} and following section). For
reference, the conversion between flux density (Jansky,
Jy)\footnote{$1\, \mathrm{Jy} = 10^{-26}\; \mathrm{W\,Hz^{-1}\,m^{-2}}
  = 10^{-23}\; \mathrm{erg\,s^{-1}\, Hz^{-1}\,cm^{-2}}$} and
brightness temperature (in Kelvin, K) is given by $\Snu=\Bnu \Omega=2
k c^{-2}\nu^2\Tb \Omega$ and hence we have for an extended source
\begin{equation}
S_\nu=0.93\,{\rm kJy}\,{\Tb\over{\rm 10^6 K}}{\Omega\over{\rm
deg}^2}\left({\nu\over{\rm 10\,MHz}}\right)^2
\end{equation}
where $\Bnu$ is the surface brightness, $\Tb$ is the surface brightness
temperature, $\Snu$ is the flux density, $\Omega$ is the source's
solid angle, and $k$ is the Boltzmann constant $1.38\times
10^{-23}$\,J/K.

Because of the absorption, the surface brightness distribution of the
Galaxy varies with frequency: at high frequencies, the Galactic plane
is seen in emission, with HII regions being superimposed in
absorption, while the poles are more or less transparent; at
frequencies around 2-3~MHz, the surface brightness is constant in all
directions, so that the sky appears equally foggy everywhere; at lower
frequencies, the plane is seen in absorption, so that the poles will
appear warmer than the plane.

\begin{figure*}
\begin{center}
\includegraphics[width=0.55\textwidth]{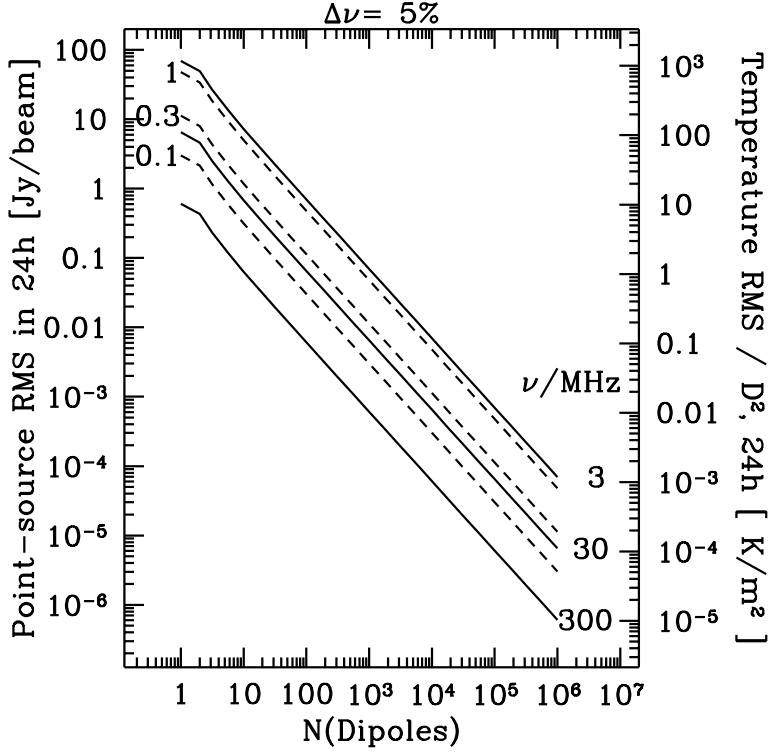}
\end{center}
\caption{\label{f:sensNant}RMS sensitivity (1-sigma noise level) of an
  array of crossed dipoles (antenna effective area = $\lambda^2/4$),
  for 5\% fractional bandwidth and 24\,h integration time, as function
  of the number of crossed dipoles in the array as given by
  eqns.~(\protect\ref{eq:arrayRMSflux}) and
  (\protect\ref{eq:dipoleTRMS}); see eq.\ (\protect\ref{eq:Nmax})
  for the maximum useful number of dipoles due to shadowing. The
  system temperature is assumed to be equal to the Galactic sky's
  brightness temperature in the direction of the Galactic poles
  (Fig.\,\protect\ref{f:galBG}). The left-hand axis gives the point-source
  sensitivity. The right-hand axis yields the surface brightness
  sensitivity in Kelvin for one resolution element after multiplying
  the given value by the square of the array's maximum baseline $D$,
  where $D$ is measured in meters. The RMS scales as the inverse of
  the square root of both bandwidth and integration time.}
\end{figure*}

The main effect of the Galactic synchrotron emission at low
frequencies is that the sky temperature is so high that it dominates
the system temperature and therefore limits the sensitivity. Figure
\ref{f:sensNant} shows the expected sensitivity for a lunar array as
function of the number of crossed dipoles, each with effective area
\begin{equation}
\Aeff = \lambda^2/4,
\label{eq:dipoleAeff}
\end{equation}
where effective area is related to primary beam size $\Omega$, which
is also the size of the antenna's field of view, by
\begin{equation}
\Aeff = \lambda^2 /\Omega.
\label{eq:Aeffdef}
\end{equation}
We use the RMS sensitivity ($1 \sigma$) of an array of antennas
\citep[][the origin of this expression is the \emph{system equivalent
    flux density}, SEFD, which is the surface brightness corresponding
  to the system temperature]{TMS01} as given by \citet{Cohen04}:
\begin{equation}
\sigma_{\mathrm{RMS}} = \mathrm{58.9\;mJy/beam} \times \frac{\Tsys}{\Aeff \sqrt{N
    (N-1)\; N_{\mathrm{IF}}\, \texp \,\Delta \nu}},
\label{eq:arrayRMSflux}
\end{equation}
where \Aeff\ is the antenna's effective area in m$^2$, $\Delta \nu$
and $\texp$ are the bandwidth in MHz and observing time in hours,
respectively, over which the signal is integrated, and $N$ is the
number of array elements (for a single dipole, the term $N(N-1)$ needs
to be replaced by 1). $\Tsys$ is the system temperature, i.e., the sky
temperature (given in Fig.~\ref{f:galBG}) and approximated here as
\begin{equation}
\Tsky = \left\{
\begin{array}{lcl}
16.3\times10^6\,K\, \left(\frac{\nu}{2\,\mathrm{MHz}}\right)^{-2.53} &\mathrm{\ at
  \ }&\nu>2\,\mathrm{MHz}\\
16.3\times10^6\,K\,\left(\frac{\nu}{2\,\mathrm{MHz}}\right)^{-0.3}&\mathrm{\ at
   \ }&\nu\leq 2\,\mathrm{MHz}.
\end{array}\right.
\label{eq:Tskyapprox}
\end{equation}
For a dish antenna, $N_{\mathrm{IF}}$ is the number of intermediate
frequency (IF) bands, i.e., 1 or 2 depending on whether only one or
both polarizations are recorded simultaneously; for our crossed
dipoles, the effective area already includes the factor of 2 for
recording both polarizations simultaneously, so that
$N_{\mathrm{IF}}=1$ is appropriate here.

For later reference, we can also cast eqn.~\ref{eq:arrayRMSflux} in
terms of surface brightness sensitivity:
\begin{eqnarray}
\TRMS &=&  \frac{\Dmax^2\,\Tsys}{\Aeff\sqrt{N
    (N-1)\; N_{\mathrm{IF}} \texp \Delta \nu}} \label{eq:dipoleTRMS}\\
&=& f \frac{\Tsys}{\sqrt{N_{\mathrm{IF}}\;\texp \Delta \nu}},
\label{eq:dipoleTRMSff}
\end{eqnarray}
where $\Dmax$ is the maximum baseline of the array setting the maximum
angular resolution and
\begin{equation}
f = \frac{\Aeff \sqrt{N(N-1)}}{\Dmax^2}
\label{eq:ffactor}  
\end{equation}
is the \emph{filling factor} of the array, i.e., roughly the fraction
of the arrays's physical area that is filled with effective antenna
area.

There is a fundamental maximum to the useful number of dipoles within a
given area: an array with resolution $\vartheta$ arcminutes can have
at most
\begin{equation}
N_{\mathrm{max}} = 4.7\times 10^{7} \left(\frac{\vartheta}{1\arcmin}\right)^{-2}
\label{eq:Nmax}
\end{equation}
dipoles before the filling factor exceeds unity, i.e., the dipoles
start becoming mutually coupled; the maximum baseline necessary to
reach this resolution is given by
\begin{equation}
\Dmax=3438\,\lambda\;\left(\frac{\vartheta}{1\arcmin}\right)^{-1}.
\end{equation}
We consider a related issue in \S\ref{s:constr.conf} below: the
maximum number of useful antennas before confusion noise (rather than
random noise) limits the array performance.

\subsubsection{Free-free absorption}
\label{s:constr.gal.ff}

The magnitude of the free-free (or thermal bremsstrahlung) absorption
coefficient $\kappa_\nu$ depends on both electron density and
temperature as
\begin{equation}
\kappa_\nu =
1.78\;\gff\;\frac{n_e^2}{\nu^2\,T_e^{3/2}}\;\mathrm{m}^{-1}
\label{eq:kappa_ff}
\end{equation}
where $n_e$ is the density in cm$^{-3}$ and $T_e$ the temperature in
$K$ of the electron distribution, $\nu$ is the observing frequency in
Hz, and $\gff$ is the free-free Gaunt factor
\begin{equation}
\gff = 10.6 + 1.6 \log T_e - 1.26\log \nu
\end{equation}
\citep{PW02}. For the parameters of the interstellar medium $n_e
\approx 0.025$\,cm$^{-3}$ and $T_e \approx 7000$\,K \citep{PW02}, the
distance at which the ISM optical density satisfies $\tau=1$ is about
1.5\,kiloparsec (kpc) at 1~MHz; this distance scales as the square of the
frequency.  For comparison, the thickness of the Galactic disc, in
which the free electrons are found, is about 1~kpc, and the diameter
of this disc is about 25~kpc. Thus, extragalactic observations are
impossible below about\,3 MHz unless there are low-density
patches. Indeed, the Galactic synchrotron spectrum
(Fig.~\ref{f:galBG}) turns over at frequencies near 3\,MHz due to
absorption (this turnover is one of the ways in which the Galactic
electron density and temperature are constrained observationally).
Variations of the electron temperature and density in the ISM will
already cause difficulties in interpreting data in the range 3-7~MHz,
since variations in the surface brightness arising from fluctuations
in the foreground emission and absorption have to be disentangled from
structures in the sources. For some kinds of sources, this may be done
by comparing the source spectra as well as their morphology, but for
extragalactic synchrotron sources with similar spectra to the Galactic
synchrotron emission, this task will be considerably more difficult
the more extended they are.

On the other hand, the absorption enables a tomography of the ISM,
since observing at different frequencies picks out absorbing
structures at different distances. This is described in
section~\ref{s:scidrv.galaxy.ism}.

\subsubsection{Brightness temperature and dynamic range}
\label{s:constr.gal.Tsys}

Below about 100 MHz, the brightness temperature of the Galactic
emission is much higher than that of any noise source in the telescope
itself and therefore sets the system temperature, $\Tsys$. Because of
the structure of Galactic emission and absorption, $\Tsys$ at any
given frequency will vary strongly with direction (for example, below
30~MHz, HII regions will be seen in absorption against the Galactic
plane). \citet{Erickson99} noted that this can lead to dynamic range
and calibration problems already when observing with the VLA's
steerable antennas at 74~MHz; these difficulties are likely to be
exacerbated by the use of lower-frequency dipoles with larger primary
beams. The operation of LOFAR will provide experience and expertise in
the calibration of an interferometer in the presence of strongly
variable backgrounds. For some applications a dynamic range of $>10^5$
is sought.  Since the field of view of dipole antennas is practically
the entire sky, the necessary dynamic range can be estimated by
comparing the flux densities of the brightest radio sources in the
sky, the supernova remnant Cas~A (diameter $\approx 6\arcmin$, flux
density 65\,kJy) and the radio galaxy Cyg~A \citep[diameter $\approx
  2\arcmin$, flux density 31.7\,kJy; both flux densities are taken
  from Tab.\ 2 of][]{radio-cal}, to the desired flux limit. For
extragalactic sources, the latter is the confusion limit (see
eq.\ \ref{eq:conflim} below).  At 15\,MHz and with a resolution of
2\arcmin, for example, the confusion limit is 0.14\,Jy, requiring a
dynamic range of $\approx 2\times10^5$ to image sources at the
confusion limit against the glare of Cyg~A.

\subsubsection{Scattering in the Interplanetary and Interstellar Media (IPM/ISM)}
\label{s:constr.gal.scatter}
 
Like all solar-system planets, the Earth is embedded in the
interplanetary medium (IPM): the solar wind, an extension of the solar
corona. The properties of the solar wind vary with the 11-year solar
activity cycle. The solar system itself is embedded in the ISM, as
already discussed above. Electron density fluctuations both in the ISM
and the IPM cause scattering of radio waves, leading to the phenomena
of diffractive and refractive scintillation, scatter broadening,
temporal broadening of pulses, and Faraday depolarization of linearly
polarized sources. The angular scattering and scintillation are akin
to ``seeing''. This is the blurring of optical images in ground-based
telescopes caused by density fluctuations in turbulent layers of the
Earth's atmosphere and the main motivation for optical space
telescopes and adaptive optics.

\subsubsection{Angular scattering}
\label{s:constr.gal.angscatt}

At frequencies below 30 MHz, the scattering is in the ``strong''
regime for both media \citep{CC74}, which means that there are no
intensity scintillations, and the scattering broadens the intrinsic
source size into a scattering angle that scales roughly as
$\lambda^2$. There is solar-wind scattering at all elongations from
the sun, but the scattering is smallest in the anti-solar direction
and increases for lines of sight passing closer to the sun. Similarly,
the ISM is completely surrounding the Earth so that there is some
scattering in all directions, but because of structures in the ISM,
the scattering is stronger for lines of sight passing through the
plane of the Galaxy and closer to the Galactic centre.

\begin{figure*}
\begin{center}
\includegraphics[width=0.55\textwidth]{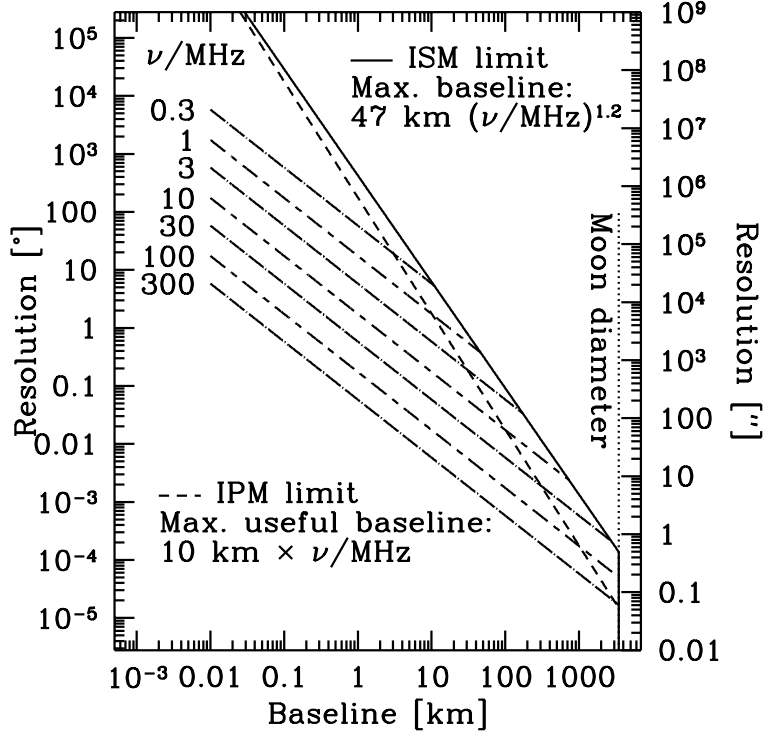}
\end{center}
\caption{\label{f:resBLscatt}Resolution of a radio array as function
  of baseline. The ISM and IPM scattering limits are also shown,
  assuming the scattering laws given by \protect\citet{Woan00}. The
  ISM scattering size is in principle a firm limit to the resolution
  achievable for extragalactic observations, but could be smaller in
  some small part of the sky, in particular away from both the
  Galactic plane and the Galactic centre. The IPM scattering has not
  been measured well at low frequencies and could be a factor 15
  smaller than given. The IPM scattering is given here for the
  anti-solar direction; it will increase when observing closer to the
  sun, and vary with the solar cycle.}
\end{figure*}
The expected order of magnitude of the scattering for both media can
be extrapolated from measurements at higher frequencies. Empirically,
the ISM broadening of the image of an extragalactic source is roughly given by
\begin{equation}
\label{eq:theta_ism}
\vartheta_{\mathrm{ISM}} \approx
\frac{30\arcmin}{(\nu/\mathrm{MHz})^{2.2}\sqrt{\sin b}},
\end{equation}
where $b$ is the Galactic latitude \citep{CC74}. Estimates of the IPM
scattering at 1~MHz range from a factor 3 smaller than the ISM
scattering \citep{RickettColes00} to a factor 5 greater
\citep{Woan00}. Figure~\ref{f:resBLscatt} shows the achievable
resolution with the limits imposed by IPM and ISM scattering, as
function of array size, using the worst-case IPM scattering  given by
\citet{Woan00}:
\begin{equation}
\label{eq:theta_ipm}
\vartheta_{\mathrm{IPM}} \approx
\frac{100\arcmin}{(\nu/\mathrm{MHz})^{2}},
\end{equation}
and a slightly different version of eq.~\ref{eq:theta_ism} for ISM
scattering, $\vartheta_{\mathrm{ISM}} = 22\arcmin
(\nu/\mathrm{MHz})^{2.2}$.  The best-case IPM scattering
\citep{RickettColes00} would be a factor of 15 smaller than given by
eq.~(\ref{eq:theta_ipm}).

The scaling $\vartheta \propto \nu^2$ arises from a thin-screen
approximation for the scattering medium.  For a screen of thickness
$L$ with overdensities of characteristic radius $a$ and excess density
$\Delta N_e$, the scattering angle is given by $\vartheta_{\mathrm{thin}}
= r_e \lambda^2 \Delta N_e \sqrt(L/a)/2\pi$, where $r_e =
2.82\times10^{-15}$\,m is the classical electron radius.  A more
realistic model of the ISM/IPM accounts for their turbulent nature,
yielding the $\nu^{2.2}$ scaling used above.  The strength of the
scattering is set by the \emph{scattering measure}, a weighted
line-of-sight integral of the turbulence power-spectrum coefficients
(see \citealp{CWFea91} and \citealp{CL02}, who also give expressions
relating the scattering measure to observables, as well as the
emission measure and the pulsar dispersion measure; \citealp{CL02}
furthermore report a model for the distribution of free electrons in
the Galaxy and give a computer programme that allows computation of
the scattering measure, angular broadening, etc.).

\subsubsection{Temporal broadening}
\label{s:constr.gal.temporal}

Together with angular broadening, electron density fluctuations due to
ISM and IPM turbulence also cause temporal broadening of radio
pulses. The temporal broadening arises because the angular broadening
allows the emission to reach the observer via multiple optical paths with
different lengths. The magnitude of temporal broadening
$\Delta\tau_{\mathrm{b}}$ is just the geometric delay between the
scattered and unscattered signal paths and therefore related to the
magnitude of angular broadening $\vartheta_{\mathrm{S}}$ by
\begin{equation}
\label{eq:delta_tau_b}
\Delta\tau_{\mathrm{b}}=\frac{z^* \vartheta_{\mathrm{S}}^2}{2c}
\end{equation}
in a small-angle approximation.  Here $z^*$ is the \emph{reduced
  distance} to the scattering screen and given by $z^* = (z
z^\prime)/(z + z^\prime)$, with $z$ the observer-screen distance and
$z^\prime$ the screen-source distance.  
For IPM scattering of Galactic
sources and for all extragalactic observations, $z^\prime \gg z$ so
that $z^* \approx z$.  \citet{Woan00} gives the following rough
scalings for temporal broadening:
\begin{eqnarray}
\mathrm{ISM: } & \Delta t & = 6\,\mathrm{yr}\;
(\nu/\mathrm{MHz})^{-4.4} \label{eq:tau_ISM}\\
\mathrm{IPM: } & \Delta t & = 0.1\,\mathrm{s}\;
(\nu/\mathrm{MHz})^{-4.4}. \label{eq:tau_IPM}
\end{eqnarray}
With these scaling laws, the IPM temporal broadening is acceptable at
least down to 1~MHz, but the ISM broadening is still 2 hours at
10~MHz, implying that no high-time resolution work can be done for
sources at extragalactic distances. For sources within the Galaxy,
$\Delta t$ increases roughly with the square of the distance to the
source for fixed turbulence strength \citep[][eq.\ 9; in addition to
  the explicit linear dependence, there is a distance dependence in
  the line-of-sight integral for the scattering measure]{CL02}; this
will be relevant for the observability of radio bursts from extrasolar
planets (\S\ref{s:scidrv.transients.extraplanets} below).  However,
there are now indications \citetext{P.~Zarka, \emph{priv.\,comm.}}
that the pulse broadening may increase less rapidly with wavelength
than expected from eq.~(\ref{eq:tau_ISM}); this might allow the
detection of ULW pulses also from extragalactic sources after all.

\subsubsection{Faraday rotation and depolarization}

The plane of polarization of a linearly polarized beam undergoes the
so-called Faraday rotation when it propagates through a region
containing both free electrons and a magnetic field component parallel
to the direction of propagation. Faraday rotation can be corrected
using multi-frequency observations and the known frequency dependence
of the effect. However, Faraday rotation for low-frequency
observations is dominated by free electrons from the solar wind in the
immediate vicinity of Earth. This makes linear polarization angles
difficult to calibrate for all sources \citep{Woan00}, since the RMS
rotation angles can be very large and time-variable.  On the positive
side this may provide some measure of solar wind activity and
distribution.

A much more severe effect is the cellular depolarization arising from
multiple uncorrelated Faraday rotation events in the ISM.  The
magnitude of the effect can be found by considering the random
walk of the polarization angle in a region with fluctuating electron
density and magnetic fields, akin to the random walk of phase in a
fluctuating electron density; assuming a constant magnetic field, the
RMS phase variation is given by 
\citep{Woan00}:
\begin{equation}
\label{eq:FaradpsiRMS}
\Delta \psi_{\mathrm{RMS}} = 2.6\times10^{-13} \; \lambda^2\;\Delta N_e \; \lambda^2 a \;
B_\parallel \sqrt{L/a} \;\mathrm{rad},
\end{equation}
where again $L$ is the path length through the scattering region, $a$
is the size of an electron density fluctuation, $\Delta N_e$ is the associated
electron overdensity, and $B_\parallel$ the magnetic flux density
component along the line of sight (all in SI units).  By inserting
into equation~(\ref{eq:FaradpsiRMS}) a size $a=10^{-3}$ parsec for
individual turbulent cells of the ISM, a typical electron density of
$0.3\times10^{5}\,\mathrm{m}^{-3}$, and a typical magnetic field of
0.5\,nT, the magnitude of the Faraday rotation at 1\,MHz is about 1
radian in one such cell. The total RMS Faraday rotation of
low-frequency radiation travelling through the ISM is therefore very
large, of order 1000 radians per parsec, and therefore leads on
average to depolarization. Thus, no linear polarization is observable
from sources beyond our own solar system at these very low
frequencies.

Circular polarization is not affected, but circularly polarized
sources are extremely rare at higher frequencies, though cyclotron
maser bursts have been observed from certain types of stars
\citep[e.g.][]{HBLea07} and is being found in an increasing number of
active galactic nuclei \citep[e.g.]{GVMO08}. At lower frequencies, cyclotron
maser bursts are expected to be visible from extrasolar planets (see
\S\ref{s:scidrv.transients.extraplanets}). Also conversion from linear
to circular emission in a dense magnetized plasma \citep[see][and
  references therein for an explanation]{BF02} is another process that can generate some
level of circularly polarized emission across a wide frequency
range. Hence, circular polarization capabilities -- which can be
measured by cross-correlating linearly polarized feeds -- should be a
more promising polarization signal to look for at low frequencies.

\subsection{Confusion and duration of a confusion-limited all-sky survey}
\label{s:constr.conf}

The first low-frequency radio surveys suffered badly from confusion:
the presence of unresolved sources with individual flux densities
below the detection limit leads to a constant floor in the noise level
that is reached after a certain observing time. This can also
lead to source positions being substantially in error
\citep[see, e.g., ][]{Kellermann03}. For low-frequency radio astronomy, a
greater level of confusion may arise from fluctuations in the Galactic
foreground emission rather than the presence of unresolved discrete
extragalactic background sources.

Among others, \citet{DMCM04} have considered the impact of these
fluctuations on the observability of the redshifted 21-cm power
spectrum at frequencies near 100~MHz. A detailed simulation of the
Galactic foreground emission based on an extrapolation of results from
WMAP and LOFAR will provide useful guidance on the expected level of
foreground fluctuations which need to be removed in order to detect
extragalactic sources. For the case of observing redshifted line
emission, spectral information will help in isolating the signal from
the foreground, and transient detection will be affected predominantly
by the bright foreground emission; however, the observability of
extragalactic continuum sources in the presence of foreground
fluctuations needs to be assessed in detail with suitable simulations.

\begin{figure*}
\begin{center}
\includegraphics[width=0.49\hsize]{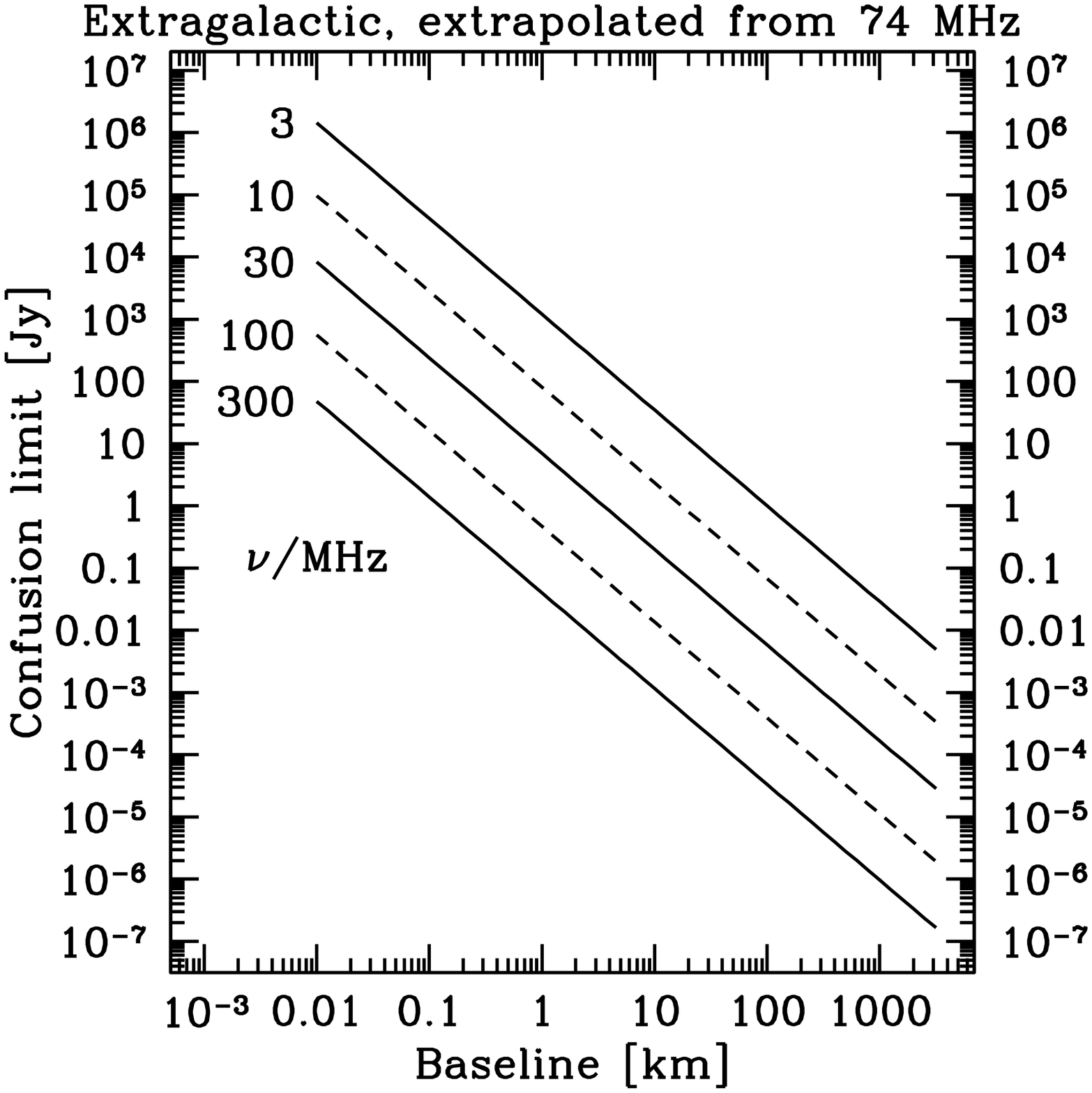}
\hfill
\includegraphics[width=0.49\hsize]{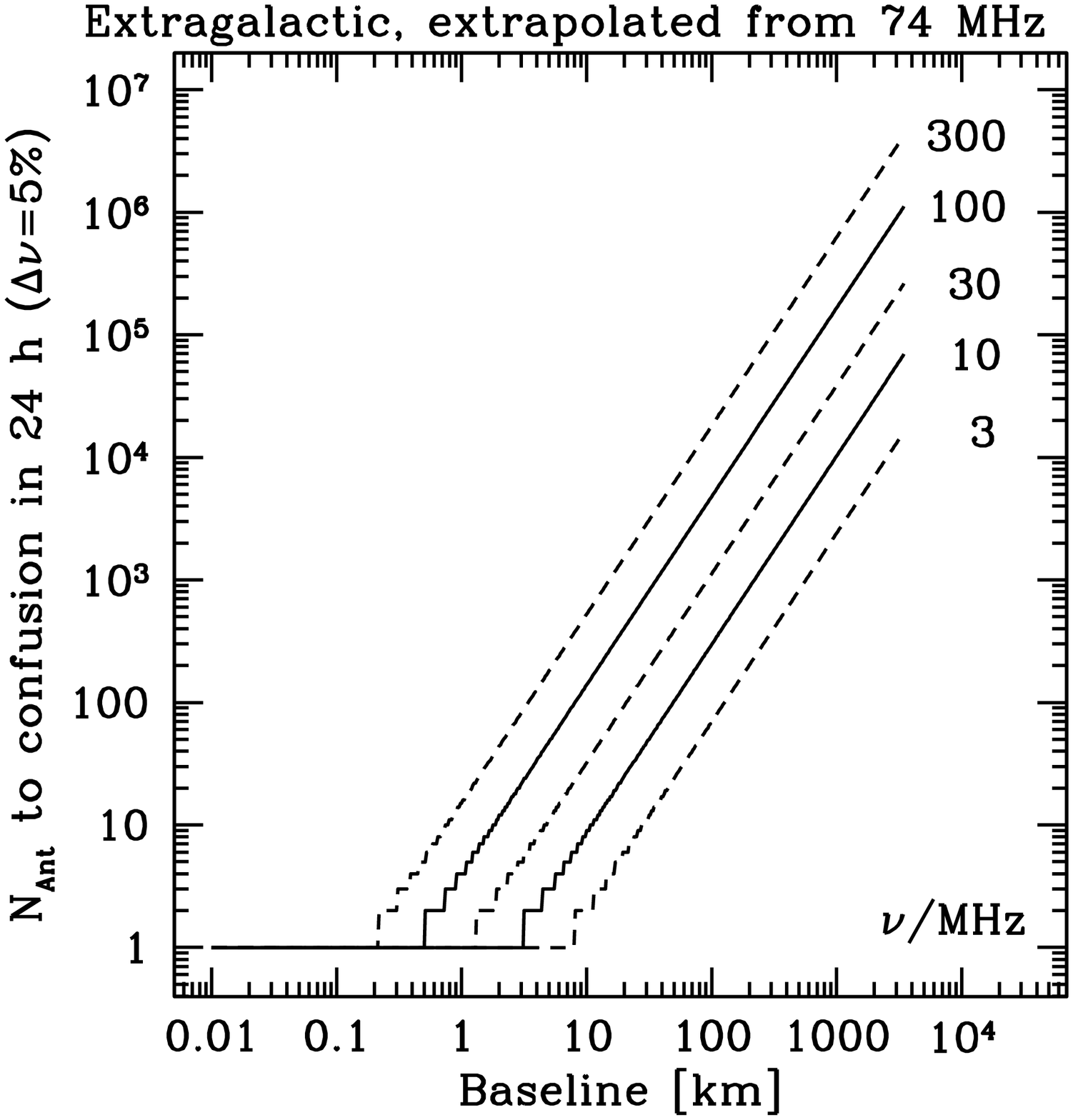}
\end{center}
\caption{\label{f:confusion}\textbf{Left:} Confusion limit from
  extragalactic background sources as function of
  baseline. \textbf{Right:} Number of antennas reaching confusion
  limit at 1-$\sigma$ level in a 24\,h integration with 5\% bandwidth;
  the integration time to reach the same limit at 5$\sigma$
  significance would be 25 times longer, or about 1 month.}
\end{figure*}
To estimate the confusion limit for extragalactic observations, which
can be done down to 3~MHz, we use the approach and formulas of
\citet{Cohen04}, who estimates the anticipated confusion limit due to
background sources by extrapolating from extragalactic source counts
and the observed confusion limit at 74~MHz.  As a conservative
estimate based on comparing the RMS noise level of 74\,MHz
observations in the VLA's C and D arrays, \citet{Cohen04} requires a
source density of less than one source per $m=12.9$ beam areas for a
survey to remain confusion-free. The beam area $\OB$ of a Gaussian
elliptical antenna beam with major and minor axes $\sigma_x$ and
$\sigma_y$, respectively, is given by
\begin{eqnarray}
\OB & = & 2 \pi\;\sigma_x \sigma_y \nonumber \\
 & \approx & 1.13\, \vartheta^2, \label{eq:beamarea}
\end{eqnarray}
where the approximation holds for a circular Gaussian beam
($\sigma_x=\sigma_y = \sigma$) of full-width at half-maximum (FWHM)
$\vartheta = \sqrt{8 \ln 2}\; \sigma \approx 2.35\; \sigma$.  The
confusion limit $S_{\mathrm{conf}}$ is then defined by
\begin{displaymath}
1.13\,\vartheta^2\, m\, N_>(S_{\mathrm{conf}}) = 1, 
\end{displaymath}
where $N_>(S)$ is the number of sources with flux densities greater
than $S$ \citep[equation 1 from][]{Cohen04}.  In the VLA Low-Frequency
Sky Survey (VLSS) that is conducted at 74\,MHz, the source counts at
the faint end are given by 
\begin{equation}
N_>(S) = 1.14\,{\rm deg}^{-2} \left({S\over{\rm Jy}}\right)^{-\beta}, \label{eq:NcountsVLSS}
\end{equation}
where $N$ is the number of sources per square degree, $\beta=1.3$, and the flux density limit $S$ is measured in Jy. To estimate $N_>(S)$ at still lower frequencies, we make the
simplifying assumption that the source flux can be extrapolated in
frequency with the spectrum $\nu^{-0.7}$. This is typical for
optically thin synchrotron sources that are expected to dominate the
extragalactic low-frequency population. Thence it follows that
\begin{equation}
N_>(S) = 1800\,{\rm deg}^{-2} \left({S\over{\rm 10\,mJy}}\right)^{-1.3}\left({\nu\over{\rm 10\,MHz}}\right)^{-0.7}. \label{eq:Ncounts10MHz}
\end{equation}
We ignore here the fact that many radio sources start showing a
low-frequency turnover due to synchrotron self-absorption, and hence
we likely overestimate the number of source and thus in turn also the
confusion limit. With this assumption, the confusion limit in Jansky
is given as function of resolution $\vartheta$ by
\begin{eqnarray}
S_{\mathrm{conf}}(\vartheta,\nu) & = &
\left(12.9\times1.14\times1.13\;\left(\vartheta/1\deg\right)^2\right)^{1/1.3} \;
\left(\frac{\nu}{74\,\mathrm{MHz}}\right)^{-0.7} \nonumber \\
& = & 16\,\mathrm{mJy} \times \left(\vartheta/1\arcmin\right)^{1.54} \;
\left(\frac{\nu}{74\,\mathrm{MHz}}\right)^{-0.7}.
\label{eq:conflim}
\end{eqnarray}
The left panel of Figure~\ref{f:confusion} shows the confusion limit
from equation~(\ref{eq:conflim}) as function of maximum baseline and
frequency (see Fig.~\ref{f:resBLscatt} for the resolution as function
of those quantities).

Since the confusion limit is a lower limit to the achievable noise, it
implies an upper limit to the useful collecting area of an array ---
adding more antennas only decreases the time in which the confusion
limit is reached, but not the overall array sensitivity.  To compute
the maximum useful number of antennas, we begin with the expression
for the RMS sensitivity of an array of dipoles,
eq.\ \ref{eq:arrayRMSflux}.  As before, we assume that each array
element is a pair of crossed dipoles with effective area given by
eq.\ (\ref{eq:dipoleAeff}).  Setting the confusion limit from
eqn.~\ref{eq:conflim} equal to the 1-$\sigma$ RMS from
eqn.~\ref{eq:arrayRMSflux} then yields the maximum useful number of
antennas for a given integration time, which is plotted in the
right-hand panel of Fig.~\ref{f:confusion} for an integration time of
24\,h.

\begin{figure*}
\begin{center}
\includegraphics[width=0.55\textwidth]{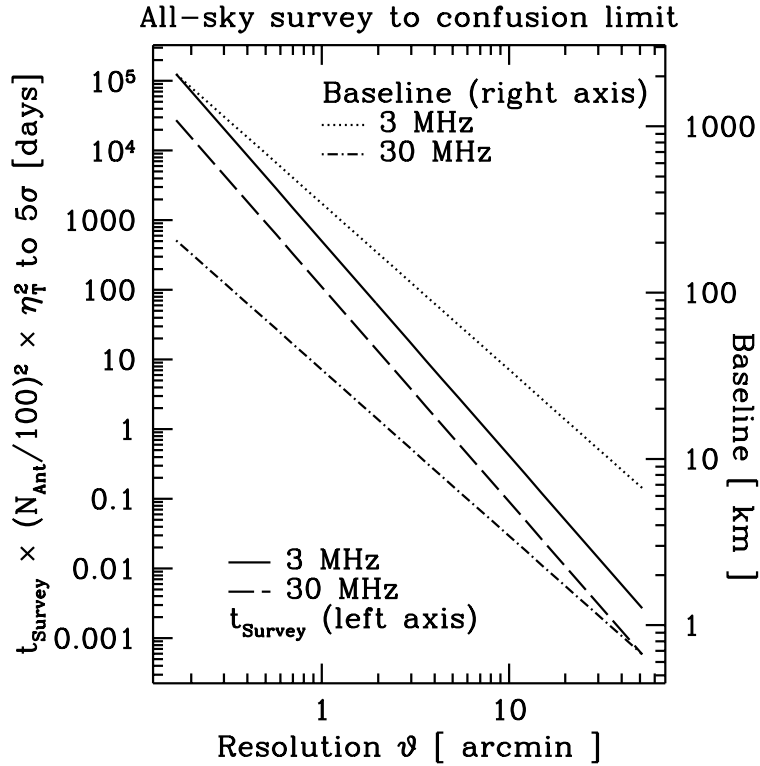}
\end{center}
\caption{\label{f:tsurvey}Exposure time (solid/long-dashed line and
  left-hand axis) and baseline (dotted/dot-dashed line and right-hand
  axis) as function of desired resolution required to perform an
  all-sky survey to the confusion limit at 5-$\sigma$ significance
  (from eq.~\protect\ref{eq:tsurvey}) with a perfectly efficient
  telescope ($\eta_{\mathrm{T}} = 1$). }
\end{figure*}
\citet{Perley01} gives an expression for the ``survey equation'', the
time needed to survey a solid angle of 1\,sr to a given sensitivity.
This is simply the time necessary for an array to reach the desired
sensitivity (via eq.\ \ref{eq:arrayRMSflux}), divided by the field of
view of each array element, i.e., the primary beam area.  For the case
of an array of $N$ crossed dipoles, with effective area of each
element given by eq.\ \ref{eq:dipoleAeff}, the time to survey one
steradian to a 1-$\sigma$ point-source flux density $s$ is given by
\begin{eqnarray}
t_\mathrm{Survey} & = & \frac{16
  \left(k\Tsys\right)^2}{s^2\,\Delta\nu N^2 \lambda^4} \nonumber \\
& = & \frac{44\,\mathrm{d}}{N^2}
\left(\frac{s}{\mathrm{Jy}}\right)^{-2} 
\left(\frac{\Tsys}{10^7\,\mathrm{K}}\right)^2 
\left(\frac{\Delta\nu}{100\,\mathrm{kHz}}\right)^{-1}
\left(\frac{\nu}{\mathrm{MHz}}\right)^{4}.
\end{eqnarray}
Inserting the high-frequency dependence of the sky temperature from
eq.~(\ref{eq:Tskyapprox}), which dominates the system temperature,
choosing the confusion limit from eq.\ (\ref{eq:conflim}) as limiting
flux density, and including a telescope efficiency $\eta_{\mathrm{T}}$,
this expression can be approximated as
\begin{equation}
t_{\mathrm{Survey}} =3.3\,\mathrm{d} \left(\frac{N}{100}\right)^{-2}
\frac{1}{\eta_{\mathrm{T}}^2}
\left(\frac{b}{0.1}\right)^{-1}
\left(\frac{\nu}{1\,\mathrm{MHz}}\right)^{-0.66} 
\left(\frac{\vartheta}{1\arcmin}\right)^{-3.08},
\label{eq:tsurvey}
\end{equation}
where $b=\Delta\nu/\nu$ is the fractional bandwidth, $N$ is the number
of antennas, and $t_{\mathrm{Survey}}$ is the time necessary to survey
one steradian of sky.  This form of the survey equation is valid at
frequencies $\nu > 2$\,MHz and as long as the sky noise temperature
over the system temperature; at frequencies below 2\,MHz, the Galactic
plane is nearly completely opaque, so that extragalactic sources
cannot be observed in any case.  An array of 300 elements and maximum
baseline 100\,km, operating at 10\,MHz (yielding a resolution of
$\approx1\arcmin$) with $b=0.1$, would thus be able to survey one
steradian of sky (3282.8\,deg$^2$) to the confusion limit in
$4/{\eta_{\mathrm{T}}^2}$ days. Note that these equations use
1-$\sigma$ sensitivities; for higher significance levels, exposure
times increase as the square of the desired significance level, e.g.,
for 5-$\sigma$ sensitivities, the required time for an all-sky survey
would be $3.5/{\eta_{\mathrm{T}}^2}$ years.  Figure~\ref{f:resBLscatt}
shows the required exposure time and baseline as function of
resolution for a confusion-limited all-sky survey with 5-$\sigma$
significance at 3\,MHz and 30\,MHz. This is roughly the range that is
accessible exclusively from above the Earth's ionosphere and not yet
optically thick to free-free scattering.

Equation~(\ref{eq:tsurvey}) gives an order-of-magnitude estimate. In
practice, details like $(u,v)$ coverage will change the actual values.
E.g., an array at a lunar pole will have more and more restricted
$(u,v)$ coverage for sources closer and closer to the lunar celestial
equator, losing the ability to resolve sources from each other in the
direction perpendicular to the equator.

As just mentioned, the presence of Galactic foreground emission adds a
new confusion component at frequencies below about 30~MHz. The
accuracy with which extragalactic source observations can be done
depends on the smallest-scale structures that are present in the
foreground emission. Observations of distant radio galaxies will be
hampered by the fact that both, they and the Galactic foreground,
produce synchrotron emission, so that the spectral signature cannot be
used for distinguishing foreground and extragalactic emission, as is
possible for 21~cm spectral line emission. As noted by Graham Woan,
``\emph{There are clearly many contributions to the apparent flux
  density of any single source, and much of the interpretation from
  discrete sources will therefore be statistical in nature}''
\citep{Woan00}. However, at least at higher frequencies most of power
in the foreground is in the diffuse component and hence higher
resolution clearly makes imaging more feasibly for any kind of survey
work.

\subsection{Stability and Calibratability}
The biggest challenge for a low-frequency phased array is the
calibration. The main issue for terrestrial arrays is calibrating
ionospheric phase fluctuations. Due to the low-density atmosphere and
ionosphere of the moon -- especially during the night -- we do not
consider this a major problem for a lunar array. This leaves
calibration of the instrumental beam shape and band pass (frequency
dependent gain of the antennas and electronics) as the main
concerns. For a phased array consisting of simple dipoles where all
beam forming and imaging happens digitally, the overall beam pattern
can in principle be calculated in a straightforward manner. However,
for terrestrial applications relative gain changes in the analog parts
of the dipole antennas need to be traced and calibrated out. For the
low-band antennas of LOFAR the main antenna gain variations are
expected to come from temperature fluctuations (affecting electronics
and damping in the wires), rain (changing the electrical properties of
ground and wires), and wind (leading to vibrations of the
antennas). Actual variations of prototype hardware in the field at
40-80 MHz have been found to be of order 20\%
\citep{NehlsLOPES07}. Since neither rain nor wind are to be expected
on the moon, temperature variations will be the largest source of gain
variations. Lunar surface temperatures can vary from $-233^\circ$\,C
to $132^\circ$\,C, well in excess of terrestrial
variations. Fortunately this change in temperature happens over the
course of a lunar day, i.e., 4 weeks and should be relatively smooth
and predictable over the array. Also, there is no variable cloud cover
that can lead to random solar irradiation. The most stable conditions
on the moon would be achieved in an eternally dark spot in a polar
crater, where one expects very little gain variations at all.

The final source of uncertainty in the overall calibration would be
the relative location of the antennas (if not known from the
deployment procedure). However, interferometric self-calibration
procedures can solve for antenna positions with high accuracy,
especially if the sky is dominated by a single (known) point
source. This is the case during Jupiter and solar bursts and has been
used effectively already to calibrate LOFAR prototype stations
\citep[such as LOPES;][]{NABea05}.

Hence, in contrast to arrays on Earth, calibratability does not seem
to be the main driver for design and layout of the array, as long as a
proper temperature control of the dipoles is achieved.

\section{Science requirements}
\label{s:scidrv}

This section describes science questions to be addressed by a lunar
array. As far as possible, the science requirements (resolution,
sensitivity) have been translated into requirements on the array
parameters (maximum baselines, number of antennas). However, there are
additional requirements with regards to the two-dimensional
distribution of antennas, the properties of the correlator, and the
methods used to calibrate the interferometric data and process them
into images of the sky. These have not been considered in detail and
are likely to modify some of the array parameters presented here. Some
of these additional aspects have been discussed by \citet{Woan97} and
\citet{JWAea98}, and other previous studies for lunar or space-based
low-frequency arrays (see citations at end of section
\ref{s:intro}).

\subsection{Cosmology with HI line emission}
\label{s:scidrv.cosmo}

A fundamental question of current cosmological research is on the
nature of structure formation in the Universe \citep[for a review,
  see][]{CiardiFerrara05}: how is the observed distribution of visible
matter created from the initial conditions just after the big bang,
when matter and radiation were distributed extremely smoothly, with
density variations of just one part in 100,000? The CMB radiation was
emitted at $z\approx 1200$, about 400,000 years after the Big Bang,
when the Universe had cooled off sufficiently for electrons and
protons to recombine into neutral hydrogen atoms (Epoch of
Recombination), allowing the background radiation to move freely
without being scattered by the electrons and protons. At the same
time, however, the Universe became opaque to visible light, because
neutral hydrogen atoms absorb visible and infrared photons and re-emit
them in random directions. Moreover, there were no sources of light in
the Universe yet: the hydrogen and helium that were created in the big
bang first had to cool in order to be able to clump together and form
stars and galaxies. Hence, this era is called the ``cosmic dark ages''
in the redshift range $z=30-1000$ \citep{Rees99}.

Things only changed after the first stars, galaxies and active black
holes had formed and emitted enough UV and X-ray photons to reionize
the neutral hydrogen, allowing all radiation to pass freely. The time
when this happened is called the Epoch of Reionization (EoR) and is
believed to have occurred around $z\sim11$, about 400 million years
after the Big Bang, though it is at present not known whether the
reionization happened more or less instantaneously, similar to a
global phase transition, or was more or less spread out in time,
depending on local conditions.

Throughout all these epochs hydrogen played a major role, emitting or
absorbing the well-known 21-cm (1.4 GHz) line due to the spin flip of
the electron. This emission is redshifted by a factor 10--1000 due to
the cosmological expansion and ends up in the frequency range from
140--1.4 MHz.  When the hydrogen spin temperature is not coupled
perfectly to the radiation temperature of the cosmic background
radiation\footnote{Which is not a \emph{microwave} background at those
  redshifts, of course.}, but changed by other couplings with the
surrounding matter and radiation, it can be seen against the cosmic
background radiation in absorption or emission, depending on whether
the spin temperature is lower or higher than the background radiation
temperature. In this way, the cosmological 21-cm emission carries
information about the evolution of the Universe.

The exploitation of cosmological 21-cm emission is subject of a
rapidly developing literature, which is reviewed by \citet{FOB06}.
Here we highlight the requirements for three applications: detecting
the global signal from the Epoch of Reionization
(\S\ref{s:scidrv.cosmo.gEoR}) as well as from the Dark Ages beyond
reionization (\S\ref{s:scidrv.cosmo.globaldark}), 21-cm tomography of
the reionization era, (\S\ref{s:scidrv.cosmo.tomo}), and measuring the
power spectrum of 21-cm fluctuations (\S\ref{s:scidrv.cosmo.powspec})
out to z=50.

\subsubsection{Global Epoch of Reionization}
\label{s:scidrv.cosmo.gEoR}

The 21-cm hyperfine transition of neutral hydrogen atoms can be used
as a cosmological probe because differences between the spin
temperature $T_{\mathrm{s}}$ characterizing the population of the
upper and lower hyperfine state and the radiation temperature
$T_{\mathrm{R}}$ of the background radiation are observable directly:
for $T_{\mathrm{s}} < T_{\mathrm{R}}$, the 21-cm emission is seen in
absorption, while it is seen in emission for the converse case
\citep{EP51}.  The spin temperature is determined by the history of
collisional and radiative excitation, which in turn is determined by
the cosmological evolution of CMB and gas temperatures \citep[see,
  e.g., ][]{SZ72,SR90}.  \citet{GS04} have simulated the evolution of
the sky-averaged brightness temperature of redshifted 21-cm emission
due to the onset of reionization, the so-called global EoR signal.
The signal is a characteristic variation of 21-cm brightness
temperature $T_{21}(z)$ with redshift, which arises at follows. Before
reionization, the hyperfine populations of neutral hydrogen are in
equilibrium with the background radiation, and $\Delta T_{21} =
T_{\mathrm{s}} - T_{\mathrm{R}}(z) = 0$.  As the first Lyman-$\alpha$
photons are generated, $T_{\mathrm{s}}$ becomes decoupled from the
background radiation temperature and couples to the gas temperature,
either by collisions or via Wouthuysen-Field scattering
\citep{Wouthuysen52,Field59}.  Since the gas temperature is
less than the CMB temperature and therefore $\Delta T_{21} < 0$, the
21-cm line is seen in absorption.  As the gas heats up, $\Delta
T_{21}$ increases and reaches values greater than 0.  The heating
increases until reionization has proceeded so far that only very few
neutral hydrogen atoms are left, thus removing the source of the 21-cm
emission and resetting $\Delta T_{21} $ to its initial value of 0.
The observed wavelength at which this signal appears directly reveals
\emph{when} reionization occurred, and the redshift evolution of
$\Delta T_{21}(z)$ encodes the detailed physics of reionization.

\begin{figure*}
\begin{center}
\includegraphics[width=0.55\textwidth]{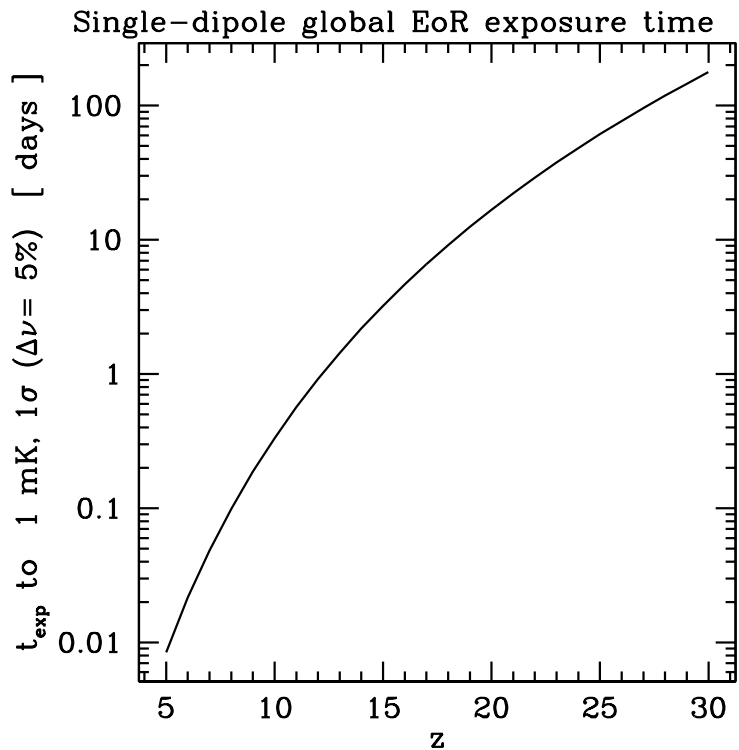}
\end{center}
\caption{\label{f:globalEoR}Integration time necessary to reach an RMS
  noise level of 1\,mK in the redshifted 21-cm line for observing the
  global Epoch of Reionization signal \protect\citep[of order of a few
    to a few tens of mK;][Fig.~1]{GS04} with a single dipole against
  the Galactic foreground emission, as function of redshift ($z=5$
  corresponds to $\nu=230\,$MHz, $z=30$ to 45\,MHz) The integration
  time is obtained from eqns.~\protect\ref{eq:Tskyapprox} and
  \protect\ref{eq:dipoleTRMSff} with $f=1$. A constant fractional
  bandwidth of 5\% was assumed.  To reach fainter noise levels $s$,
  the required integration time scales as $\texp\propto 1/s^2$.  The
  redshift range shown includes the entire likely range for the origin
  of the global EoR signal; redshifted 21-cm emission from higher
  redshifts is considered separately below.}
\end{figure*}
Because of the homogeneity and isotropy of the Universe, this
transition happened roughly simultaneously in all parts of the
Universe. Therefore, the same signal is observed independently of the
direction one is looking in, and a single simple dipole antenna would
be sufficient equipment to pick up the signal, which has an amplitude
of a few to a few tens of milliKelvin \citep[Fig.~1]{GS04}.  Thus,
although simple in principle, detection of the signal is difficult
against the Galactic foreground.  As reionization is believed to occur
in the redshift interval $6 < z < 20$, the signal will appear at
frequencies between 70 and 200~MHz.  As discussed by \citet{GS04}, the
main problem here is not the sensitivity as such. The sky temperatures
in this frequency range are 100--2000~K, with lower temperatures at
higher frequencies (eq.~\ref{eq:Tskyapprox}). As
Fig.~\ref{f:globalEoR} shows, the necessary observing times to reach
10-$\sigma$ detections of a 1~mK signal from $z=15$ with a single
dipole are just a few months. Hence, the ability to distinguish the
global EoR signal from the effect of the foreground variation imposes
a further observability constraint.  The evolution of $\Delta
T_{21}(z)$ corresponds to a smooth change of the spectral index as
function of frequency, which needs to be distinguished from similarly
smooth variations in the Galactic foreground emission.  Its detection
is -- so far unsuccessfully -- being tried from the ground and
requires very high gain stability and absence of RFI: the Australia
Telescope National Facility has an ongoing development effort
``Cosmological Reionization Experiment'' (CoRE) that attempts a global
EoR measurement \citep{CRE05}; \citet{BRH08} have just reported first
results from the EDGES
experiment\urlfn{http://www.haystack.mit.edu/ast/arrays/Edges/},
determining an upper limit of 250\,mK to the reionization signal, and
stressing the need for systematic errors being limited to less than
3\,mK.

A single antenna with an instantaneous frequency range 10-200 MHz, in
an ``eternally dark'' lunar crater at the pole, kept at a stable
temperature, and shielded from any solar and terrestrial radiation,
would clearly offer the very best set-up for such an experiment. That
would reduce most instrumental and RFI effects to the bare minimum and
only leave the cosmic foreground as the main issue. Using several
independent dipoles, without any beam-forming, would increase
sensitivity and reliability against systematic errors even
further. With enough signal paths the same antennas could then also be
used in a beam-forming mode. Systematics and cosmic variance could be
further checked by using both lunar poles which lets one observe two
independent patches of the sky.  A lunar array could also collect the
global EoR signal simultaneously with other observations by adding
signals from different antennas incoherently. This only requires
including a dedicated signal processing chain parallel to the
correlator.  Together this should deliver an ultra-precise global
radio spectrum of the low-frequency sky to identify the brightness
temperature change in the the 21~cm line caused by reionization. This
would answer the question: When did the global transition between a
predominantly neutral and a predominantly ionized Universe happen?

\subsubsection{The global 21-cm signal from the Dark Ages beyond
  reionization}
\label{s:scidrv.cosmo.globaldark}

\begin{figure*}
\begin{center}
\includegraphics[width=0.55\textwidth]{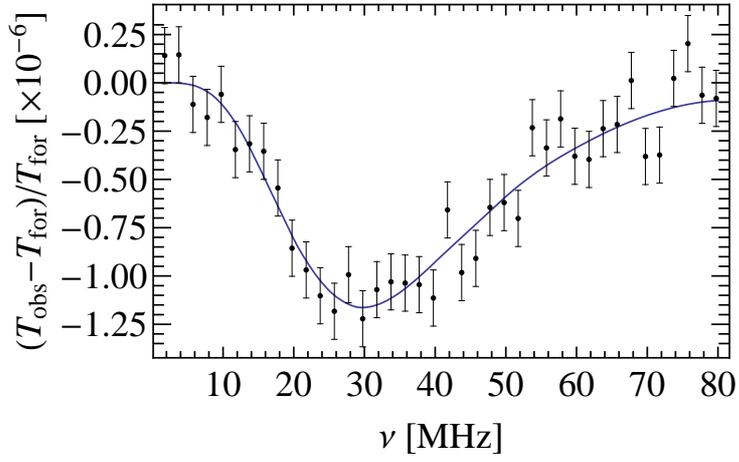}
\end{center}
\caption{\label{f:darkages-line} A simulated global 21-cm HI signal
  from the dark ages after 1 year of integration time with a single
  sky-noise limited dipole as a function of frequency \citep[based on
    the calculations by][also compare
    Fig.~\protect\ref{f:global21cm}]{CS07}. Here a fixed bandwidth of
  1.5 MHz is used. We assume that the observed brightness temperature,
  $T_{\mathrm{obs}}$, is just the sum of the dark ages signal and a
  foreground signal, $T_{\mathrm{for}}$, which is a perfect power law.
  The signal at 30\,MHz originates from $z=46$.}
\end{figure*}
\begin{figure*}
\begin{center}
\includegraphics[width=0.55\textwidth]{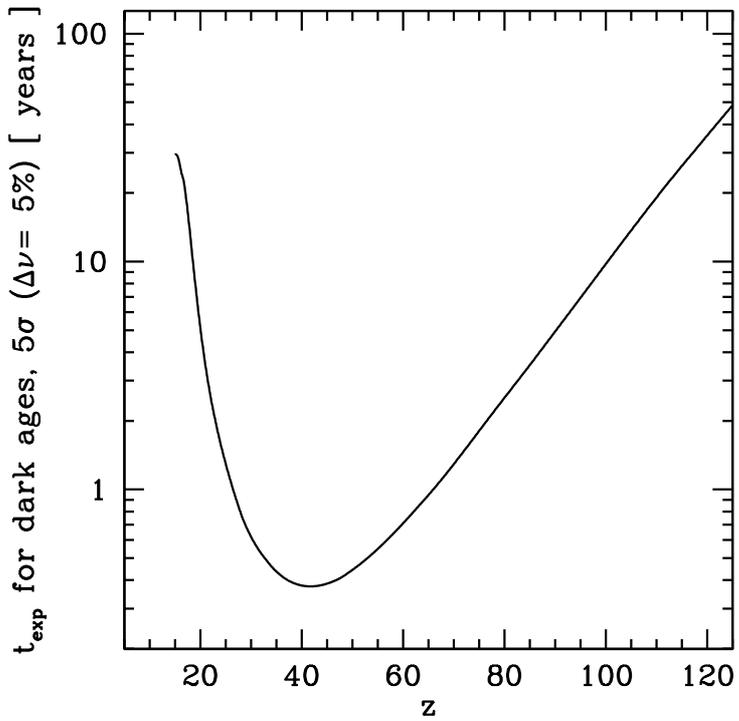}
\end{center}
\caption{\label{f:global21cm}Global 21-cm signal from HI in the dark
  ages. The plot shows the observing time needed to reach a 5-$\sigma$
  detection at 5\% fractional bandwidth with a single dipole, as
  function of redshift.  If $N$ dipoles are added incoherently, then
  the observing time is reduced by a factor $\sqrt{N}$.}
\end{figure*}

In the existing literature, the focus for 21-cm cosmology has been put
on the global signal from the Epoch of Reionization on the one hand,
and on the power spectrum of 21-cm fluctuations from the EoR and
higher redshifts on the other (see following sections). However, there
is also a global signature of the dark ages from the spin temperature
evolution at redshifts beyond reionization and up to $z\approx 200$
\citep[e.g.,][]{CM03,CS07}.  The expected signal is a surface
brightness decrement with a peak amplitude of around 40\,mK at
$z\approx90$ and approaching 0 at $z\approx 20$ and $z\approx 400$.
We have calculated the exposure time necessary to observe this
\emph{global Dark-Ages signal} with a single dipole against the sky
background, using the model of \citet{CS07}. The result is shown in
Fig.~\ref{f:darkages-line}, with the exposure time necessary to reach
a 5-$\sigma$ detection for a signal peaking at different redshifts
shown in Fig.~\ref{f:global21cm}. A single dipole can provide a
5-$\sigma$ detection at $z=42$ in 137 days.  A caveat here is that the
presence of the reionizing Lyman-$\alpha$ photons may be tightly
correlated with high densities also beyond the very earliest stages of
reionization.  In this case, the final spin temperature may be much
higher than the originally predicted diffuse hydrogen temperature,
i.e., the spin temperature is closer to the CMB temperature,
decreasing the absorption dip and weakening the absorption signature.

The simulated line detection with Gaussian noise in
Fig.~\ref{f:darkages-line} shows that the signal is only $10^{-6}$ of
the foreground signal. Moreover, a detection has to rely on the
foreground to be a perfect power law. Hence, the foreground
subtraction is the biggest uncertainty together with the need for an
almost perfect band-pass calibration of the antenna.  It is likely
that the actual foreground signal will deviate from a simple
power-law. This makes a detection almost impossible if there are
spectral foreground fluctuations of the same magnitude as the dark
ages signal. Nonetheless, further studies need to show on which level
spectral foreground fluctuations are actually present.  As in the
global EoR case, the global dark-ages 21-cm signal is isotropic, so
that a single (well-calibrated) dipole suffices to carry out this
measurement and detect the signal from Fig.~\ref{f:darkages-line} in
frequency space.

\subsubsection{EoR tomography}
\label{s:scidrv.cosmo.tomo}

\begin{figure*}
\begin{center}
\includegraphics[width=0.55\textwidth]{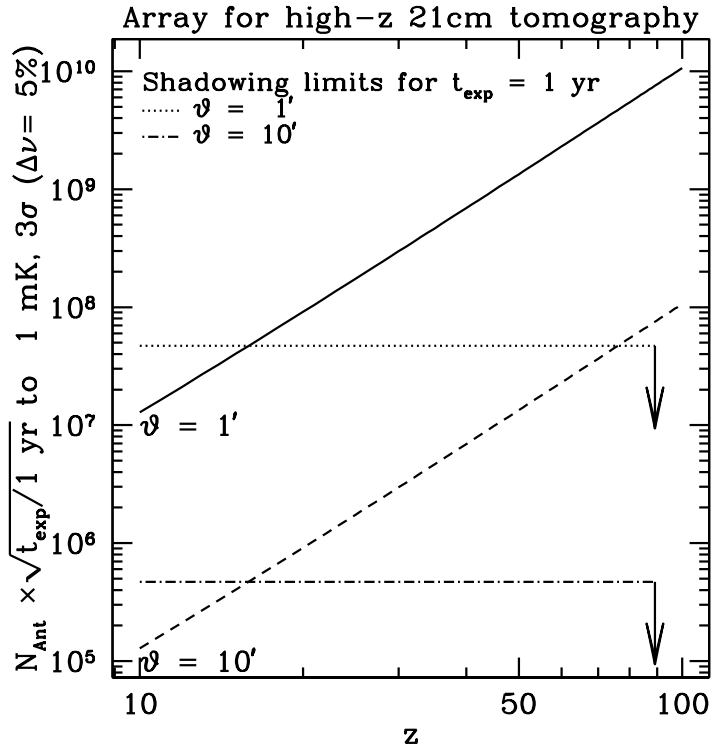}
\end{center}
\caption{\label{f:EoRtomography} Number of crossed dipoles (from
  eq.~[\protect\ref{eq:arrayRMSflux}]) required for a 3-$\sigma$
  detection of 1~mK per resolution element in one year when imaging
  the redshifted 21cm emission at frequency
  $\nu_{21}(z)=1.4\,\textrm{GHz}/(1+z)$ at angular resolution
  $\vartheta=1\arcmin$, assuming a constant filling factor independent
  of wavelength (i.e., a scaled array). The required number of crossed
  dipoles to achieve this RMS can be traded against longer exposure
  times (and \emph{vice versa}) as indicated, but must not exceed the
  shadowing limit from eq.\ (\protect\ref{eq:Nmax}),
  $N<N_{\mathrm{max}} = 4.7\times 10^{7}
  (\vartheta/1\arcmin)^{-2}$. The corresponding collecting area is
  given by $1.1\times10^{-8}\,N(1+z)^2\,{\rm km}^2$. The array
  diameter (or maximum baseline) required to reach a given resolution
  is shown in Fig.\protect\ref{f:resBLscatt} and ranges from $\approx
  800m$ ($z=10,\vartheta=10\arcmin$) to $\approx 80\,km$
  ($z=100,\vartheta=1\arcmin$). The numbers from this graph can be
  scaled as $N_{\mathrm{Ant}}\sqrt{t_{\mathrm{exp}}}\, S^{-1}\,
  \vartheta^{2} (1+z)^3 = \mathit{const}$ (where $S$ is the desired
  sensitivity).}
\end{figure*}

At the other extreme of the spectrum of potential cosmology
experiments would be an attempt to image the hydrogen fluctuations
directly. The detailed history of reionization can be traced by
observing the two-dimensional structure of neutral and reionized gas
around luminous objects, and by stepping through the different
frequencies corresponding to different emission redshifts, a
tomographic map of reionization can be created. The characteristic
scale of the fluctuations is expected at the arcminute scale
\citep{FSH04}. This adds arcminute-scale spatial resolution over the
frequency range 50--140\,MHz (corresponding to redshifts $10< z < 20$)
to the requirements, as well as the ability to detect cosmological
milli-Kelvin brightness fluctuations over the scale of one resolution
element.  The same hardware would allow to image hydrogen
fluctuations directly also during the preceding dark ages, at $30 < z
< 100$, if the frequency range is extended down to 14\,MHz and the
baselines are increased correspondingly to achieve the desired
resolution also at these longer wavelengths.

The purity requirements for EoR likely mandate a lunar far-side
location to ensure absence of solar, terrestrial auroral, and man-made
interference. A near-polar region could be considered since this will
provide a near-constant view of the sky, allowing long
integrations. If the site is closer than $\sim$6.5$^\circ$ to the
pole, lunar libration may expose the site to the Earth and make EoR
observations impossible for some fraction of the year.

The expression for the exposure time necessary to reach a given
surface brightness sensitivity \TRMS\ is particularly simple when
expressed as function of the system temperature (which is a function
of observing frequency, i.e., redshift, via
$\nu_{21}(z)=1.4\,\mathrm{GHz}/[1+z]$) and given by solving
eq.~(\ref{eq:dipoleTRMSff}) for the exposure time:
\begin{eqnarray}
t_{\mathrm{exp}} & = & \frac{1}{\Delta \nu}
\left(\frac{\Tsky\!\left(\nu_{21}[z]\right)}{f\;\TRMS}\right)^2
\nonumber \\ &=& 
\frac{2.1\times10^{12}\,\mathrm{yr}}{N(N-1)} \left(\frac{b}{5\%}\right)^{-1}
\left(\frac{\TRMS}{1\,\mathrm{mK}}\right)^{-2}
\left(\frac{\vartheta}{1\arcmin}\right)^{-4} \left(\frac{1+z}{11}\right)^{6.06}
\label{eq:texp_EoRtomo}
\end{eqnarray}
where $f$ is the usual filling factor (eq.~\ref{eq:ffactor}) and $b =
\Delta\nu/\nu$ is the fractional bandwidth.  Here, we have used the
fact that the sky temperature happens to take on the simple form
\begin{equation}
  \Tsky\!\left(\nu_{21}[z]\right) = (1+z)^{2.53}\times 1\,\mathrm{K}
\label{eq:Tsky21}
\end{equation}
using the approximation to $\Tsky(\nu)$ from eq.~(\ref{eq:Tskyapprox})
for the observing frequencies $\nu>2$\,MHz considered here, and
substituting $\nu_{21}[z] = 1.4\,\mathrm{GHz}/(1+z)$ (see also upper
panel of Fig.~\ref{f:pspecObs}).

The exposure time from eq.~(\ref{eq:texp_EoRtomo}) is plotted against
target redshift in Fig.~\ref{f:EoRtomography}, assuming a scaled array
in which the maximum baseline increases with wavelength to keep the
resolution constant at $\vartheta=1\arcmin$. This implies a filling
factor $f=2.1\times10^{-8} \sqrt{N(N-1)}$; conversely, the number of
dipoles required for a given filling factor at a given resolution is
\begin{equation}\label{eq:ndipeorim}
N_{\mathrm{dipoles}} \approx 47\times10^6\;
f\;\left(\frac{\vartheta}{1\arcmin}\right)^{-2}
\end{equation}
for $N_{\mathrm{dipoles}} \gg 1$. The corresponding effective area is
$N_{\mathrm{dipoles}}\times \lambda^2/4$, i.e.,
\begin{equation}
A_{\mathrm{tot}}=10.5\,\mathrm{km}^2 \;N_{\mathrm{dipoles}}
\left(\frac{\lambda}{30\,\mathrm{m}}\right)^2
\;\left(\frac{\vartheta}{1\arcmin}\right)^{-2},
\label{eq:Array_Aeff}
\end{equation}
where $1\,\mathrm{MHz}/\nu$ can be substituted for the
$\lambda/30\,\mathrm{m}$ term.  

Again, the exposure time is driven by the large brightness of the
Galactic synchrotron emission, which sets the system temperature.  The
exposure time can only be decreased by increasing the filling factor,
i.e., the number of dipoles; since the filling factor is limited to
$f\leq 1$, setting $f=1$ in eq.~(\ref{eq:texp_EoRtomo}) yields a firm
lower limit to the necessary exposure time for a given $\TRMS$.  In
principle, if enough correlator capacity is available, the imaging
could be performed over a significant fraction of the entire visible
sky within that time scale, since dipole antennas have a very large
instantaneous field-of-view (compare the discussion of survey speed in
\S\ref{s:constr.conf}).

It is evident that an array with a substantial number of dipoles is
required to reach the sensitivity necessary for 21-cm tomographic
imaging. For observing times of order one year for imaging around
$z=20$ (at 66 MHz or 4.4~m), of order $10^7$ dipoles would be
required, with a total effective area of 50\,km$^2$ (also see
Fig.~\protect\ref{f:sensNant}). The strongest factor here is the
spatial resolution which enters with the fourth power in the
integration time (eq.~\ref{eq:texp_EoRtomo}) and squared in the number
of dipoles needed (eq.~\ref{eq:ndipeorim}). Hence, a 10\arcmin\ array
observing 21-cm emission from $z=20$ \citep{ZLMea07} would require
only about $10^5$ dipoles (i.e. 0.5\,km$^2$) at the same sensitivity
limit. Higher redshifts -- in the pre-EoR area -- become increasingly
more difficult even though the collecting area per dipole increases
with $\lambda^2$ (i.e., $(1+z)^2$) because the sky temperature
increases rapidly with longer
wavelengths. Fig.~\protect\ref{f:sensNant} and
eq.~(\ref{eq:texp_EoRtomo}) imply that at the highest redshifts the
number of dipoles needed would require a filling factor larger than
unity for one year of integration time and 1\arcmin\ resolution. This
is of course impossible and can only be remedied by using antennas
with a higher gain than the crossed dipoles assumed here. For example,
log-periodic or Yagi antennas have a higher gain (i.e., effective
area) per antenna, at the cost of a smaller field-of-view and, worse,
increased complexity compared to simple crossed dipoles. As
Fig.~\ref{f:EoRtomography} shows, a filling factor $f=1$ is already
needed to achieve exposure times of order 1\,year for tomography at
$z\approx 15$; since the exposure time at fixed resolution and filling
factor grows roughly as $(1+z)^6$, substantially higher redshifts can
only be reached with very high-gain antennas or very long observing
times.

Confusion noise poses further constraints: 21-cm emission from $z=15$
is observed at 87.5\,MHz, where the confusion noise at
1\arcmin\ resolution is about 14\,mJy, which corresponds to a
brightness temperature of 710\,K.  This is much more than the desired
signal of 1\,mK.  To resolve the confusing background and reduce it to
the 1\,mK level, which is prerequisite for allowing its subtraction,
it would be necessary to improve the resolution to the 0\farcs01
level, which would require baselines larger than the diameter of the
moon (3476\,km).  Instead, again statistical techniques have to be
employed to extract the signal from the confusion noise via their
different structures in frequency space \citep{DMCM04}.

In summary, one can conclude that a lunar EoR imaging experiment needs
to be very substantial in size to penetrate into the $z>20$ epoch, but
is not unthinkable in the long run. This crucially depends in the
expected angular size scale of the signal, which will hopefully become
more clear with advances in ground based experiments.  Accurate
subtraction of the sources of extragalactic confusion noise is also a
necessity and may prove difficult.

\subsubsection{Power spectrum of the 21cm-transition at $z=30$--$50$}
\label{s:scidrv.cosmo.powspec}

\citet{LZ04} have pointed out that the fluctuation power spectrum of
the redshifted 21-cm line carries a wealth of information about the
matter power spectrum. The CMB radiation itself that is observed today
carries information about cosmological parameters mainly at the
largest angular scales from 0\fdg2--90\deg\ (multipole scale $l$ from
2--1000, and limited in principle by Silk damping to $l < 3000$). By
contrast, the angular power spectrum in the redshifted 21-cm line
carries cosmological information at much smaller angular scales, $l\ga
10^4$, corresponding to angular scales of 1\arcmin\ or less. In
addition, observations of redshifted 21-cm emission from different
redshifts in the range 30--50 yield independent samples of the
cosmological parameters, while the CMB information suffers from cosmic
variance. Therefore, observations of the 21-cm power spectrum from
this redshift interval yield many orders of magnitude more information
about density fluctuations in the early Universe than direct
observations of the CMB power spectrum. This makes the redshifted
21-cm line a powerful tool to constrain \emph{all} model parameters
necessary to describe the Universe, e.g., the slope and curvature of
the initial density fluctuation spectrum containing information from
the inflationary phase, the mass of warm dark matter particles, and
the fraction of matter density contributed by neutrinos.

In particular, angular scales $l > 5000$ carry the greatest weight in
discriminating between different values for the cosmological
parameters, since it is at these wavenumbers where the predicted power
spectra for different sets of cosmological parameters differ most
\citep[Fig.~3]{LZ04}. A low-frequency measurement of the 21-cm power
spectrum at these angular scales and in this redshift interval would
therefore be an ultimate cosmological experiment.

Baryon acoustic oscillations (BAOs) are a major tool of modern
cosmology, producing a feature that is directly visible in an observed
power spectrum. They are imprinted on the 21-cm angular power spectrum
in the same way as on the CMB \citep{deBernardisea00} and galaxy power
spectra \citep{Eisensteinea05,Coleea05,Percivalea07}.  The BAO signal
is present on spatial scales of 20--400\,Mpc (comoving;
\citealp{BL05}, Fig.~4).  In the 21-cm angular power spectrum, it is
predicted to appear at wavenumbers of order $l \approx 90$--4000 and
with oscillation amplitudes of order or milliKelvin or smaller (see,
e.g., \citealp{LC07}, Fig.~8 and \citealp{MaoWu08}, Fig. 4).

The distinct advantage of observing the 21-cm power spectrum from the
Dark Ages, before the Epoch of Reionization, is that the astrophysics
of reionization is complex, non-linear and not well understood, but
these complexities do not affect the Dark-Ages power spectra, which
therefore provide a much cleaner signal to observe and to
interpret. For example, the non-Gaussianity of the fluctuations can be
probed \citep{CoorayLiMelchiorri08} and may reveal information on the
inflationary phase of the universe.

\begin{figure*}
\begin{center}
\includegraphics[width=0.55\textwidth]{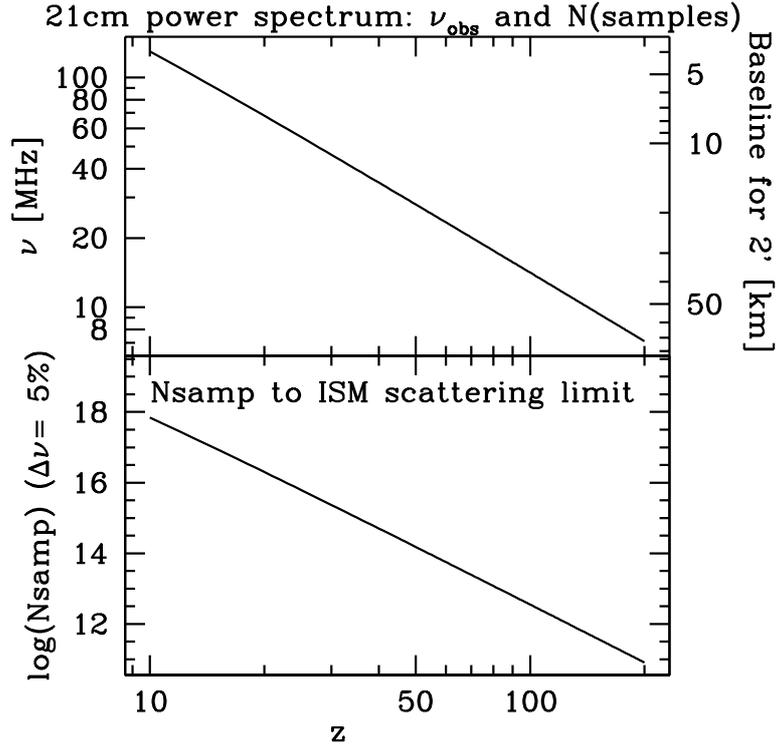}
\end{center}
\caption{\label{f:pspecObs}\textbf{Upper panel:} observed frequency of
  redshifted 21cm emission (left-hand axis), and baseline required to
  reach 2\arcmin\ resolution at that frequency (right-hand axis), both
  as function of redshift $z$. Ground-based observations are
  exceedingly difficult below 30 MHz, corresponding to
  $z=47$. \textbf{Lower panel:} Number of independent power-spectrum
  samples \protect\citep[$N_{\mathrm{21cm}}$ from][p.\ 4]{LZ04}
  obtainable down to the ISM scattering limit towards the Galactic
  poles, also as function of redshift. }
\end{figure*}
\begin{figure*}
\begin{center}
\includegraphics[width=0.55\textwidth]{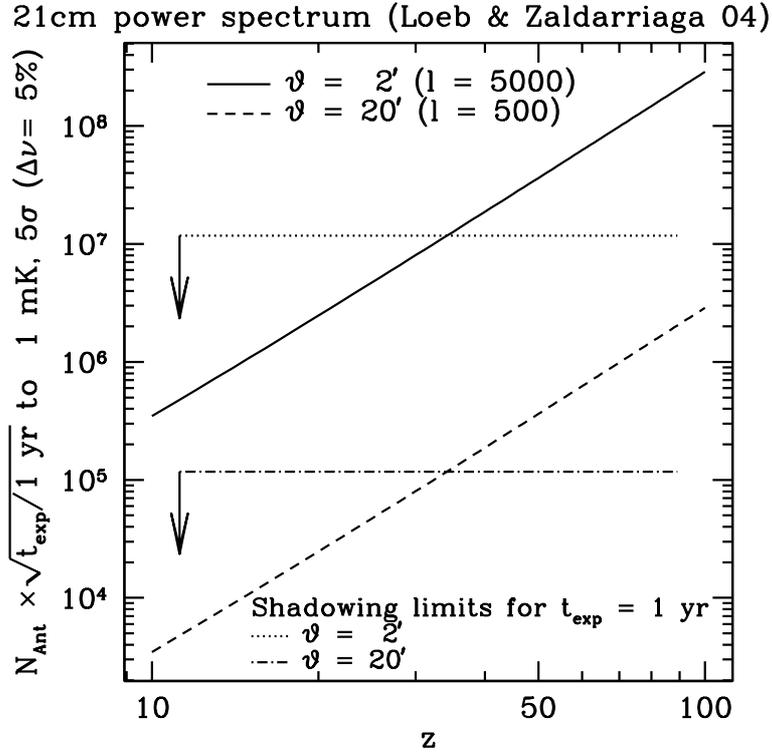}
\end{center}
\caption{\label{f:LZ04powspec} Number of antennas necessary to achieve
  a 5-$\sigma$ detection of fluctuations in the 21\,cm power spectrum
  in one year at multipole numbers up to $l=5000$ (corresponding to
  2\arcmin\ angular resolution) or $l=500$ (20\arcmin) at the 1~mK
  level, as function of redshift \protect\citep[as given by eq.~11
    from][]{LZ04}.  Except for the different exposure time
  prescription, this figure is constructed in the same way as
  Fig.~\protect\ref{f:EoRtomography}, so that the same scalings apply,
  \emph{i.e.}, $N_{\mathrm{Ant}}\sqrt{t_{\mathrm{exp}}}\, S^{-1}\,
  l^{-2} (1+z)^3 = \mathit{const}$ (where $S$ is the desired
  sensitivity level), assuming a filling factor independent of
  frequency.}
\end{figure*}
The required observing frequency is 1.4 GHz$/(1+z)$; see the upper
panel of Figure~\ref{f:pspecObs}. As noted in section
\ref{s:constr.gal.angscatt}, angular scatter broadening in the ISM and
IPM limits the highest observable angular scale as a function of
frequency. The limiting angular scale increases proportionally to
$\lambda^2$ (see Fig.~\ref{f:resBLscatt} for the maximum possible
resolution and the baseline necessary for reaching this
resolution). The lower panel of Figure~\ref{f:pspecObs} shows the
number of independent samples (data points) that can be collected up
to the IPM scattering limit \citep[$N_{\mathrm{21cm}}$
  from][p.\ 4]{LZ04}, assuming a bandwidth of 5\% of the observing
frequency. Since the number of samples scales as $\lmax^3$ and $\lmax$
scales as $\lambda^{-2}$, which in turn scales as $(1+z)^{-2}$, the
number of samples up to the scattering limit scales as
$(1+z)^{-6}$. Therefore, observing fluctuations originating from
$z=45$ only contributes 10\% of the information accessible at z=30,
and observing $z=70$ only contributes an additional 1\% compared to
$z=30$. Thus, as far as the number of samples is concerned, the
presence of angular broadening in the ISM and IPM makes extending this
experiment to lower frequencies progressively more
inefficient. However, the detailed strength of the signal as function
of redshift cannot be predicted in advance, and there may be highly
relevant deviations from current theory. This is particularly
important since, as \citet{LZ04} point out, this phase of the
cosmological evolution is a fairly clean laboratory, hardly affected
by any 'dirty' astrophysics of the reionization process. Measuring the
evolution of structure formation with redshift between $z=10-100$ may
therefore be one of the most intriguing diagnostics of the history of
the Universe.  For these reasons, it will still be desirable to cover
a broad redshift, i.e. frequency, range.

Sensitivity poses another constraint. The angular power spectrum has
the strongest total signal at $z\approx 55$ (25 MHz), dropping off
roughly symmetrically to $z=25$ and $z=170$ \citep[Fig.~2,
  e.g.]{LZ04}. The signal strength is expected to be at the mK scale,
while the system temperature and hence the noise are determined by the
sky temperature (see Figure~\ref{f:galBG}). Figure \ref{f:LZ04powspec}
shows the observing time necessary to reach an RMS level of 1~mK in
the power spectrum with an array of dipoles as given by
\citet[eq.~11]{LZ04}. To achieve an angular resolution of 1\arcmin\ in
reasonable integration times of order one year, about $10^{5.5}$
dipoles are needed for $z=30$ ($A_{\rm eff}=$3.5\,km$^2$) and $10^6$
dipoles ($A_{\rm eff}=$30\,km$^2$) at $z=50$. Again, this number
depends very critically on the angular resolution required and would
reduce to $N\approx10^{3.5}$ and $N\approx10^{4}$, respectively, for
angular scales of $10\arcmin$ if the signal were on the mK level.
Distinguishing between different cosmological models at high
confidence levels or detecting BAOs directly will probably require
sensitivities down to the 0.01\,mK level, but the latter may already
be feasible with degree-scale resolution, probing multipole numbers
$l\approx100$ \citep{BL05}.

Again, accurate subtraction of spectral and noise power fore- and
backgrounds is essential. Determining the exact array parameters that
will be necessary requires more detailed guidance from further
cosmological observations and simulations.  Numerous novel approaches
to extracting cosmological information from 21-cm power spectra are
currently being proposed to overcome the observational challenges
\citep[see,
  e.g.,][]{MBCea06,BarkanaLoeb08,VisbalLoebWyithe08,CoorayLiMelchiorri08}.

\subsection{Extragalactic surveys: radio galaxies, clusters and feedback}
\label{s:scidrv.extragal}

At low radio frequencies, the brightest sources of the extragalactic
sky are the synchrotron-emitting radio lobes of giant radio galaxies
and quasars, such as Cygnus A. In fact, the 178 MHz 3C survey
\citep[now used in its twice-revised form 3CRR,][]{LRL83} was the
first successful survey of discrete sources, and 3C sources are still
studied extensively today. There are a number of questions that are
currently of particular relevance to which the study of radio galaxies
can provide decisive insights.

\subsubsection{Electron energy distributions and jet power in radio galaxies} 
\label{s:scidrv.extragal.edist}

The low-frequency emission of radio galaxies and quasars is
generated predominantly in the ``lobes''. These are structures with
scales of tens to hundreds of kpc \citep{Miley80,BBR84} that are
inflated by pairs of relativistic jets emerging from near the
accretion disk around a black hole in the nucleus of the host galaxy.

Jets and lobes of radio galaxies emit synchrotron radiation, which has
the property that the lowest-frequency emission is generated by the
lowest-energy electrons. Since electron energy distributions typically
are of the form $N(E)\,\de E\sim E^{-p}\,\de E$ with $p\approx 2.4$ to
yield the observed spectra $\nu^{-0.7}$, low-energy electrons are the
most numerous and dominate the power output. On the other hand, as
discussed below, higher-energy electrons have shorter cooling time
scales, leading to a steepening of the radio spectrum -- the so-called
spectral aging \citep{1987MNRAS.225....1A,2002astro.ph..9372B}. Hence,
older parts of a source tend to have steeper radio
spectra. Low-frequency observations thus probe the oldest structures
in a galaxy and can be used to constrain the age of a source.

If a source is sufficiently compact, it can become optically thick to
its own synchrotron emission due to synchrotron self-absorption. This
changes the observed spectral energy distribution from the intrinsic
optically thin synchrotron spectrum $f_\nu \propto \nu^\alpha$, with
$\alpha \approx -0.7$ as typical value, to $f_\nu \propto \nu^{5/2}$,
censoring some of the observables by hiding spectral structures and
effectively removing access to the low-energy electrons.  The
self-absorption limited maximum brightness temperature scales as
$T_{\mathrm{max}}\approx
10^{12}\,\mathrm{K}\,(\nu/1\,\mathrm{MHz})^{1/2}\,(B/1\mathrm{nT})^{-1/2}$
\citep[eq.~A8]{BBR84}. Typical magnetic fields in lobes are estimated
to be of the order of a few nanoTesla. Lobe surface brightnesses are
well below $10^{12}$\,K, the maximum allowed by ``inverse-Compton
cooling catastrophe'' \citep{1969ApJ...155L..71K}. The lobes of radio
galaxies are typically not affected by optical-depth effects.

This is, however, not true for the ``hotspots'', the terminal shock
front of powerful jets. These frequently show a spectral turnover at
frequencies below 100 MHz. From the size of a hotspot and the turnover
frequency one can deduce reasonable limits of the magnetic field
strength and power of the jet feeding it.

The inflation of lobes by jets means that mechanical work is done on
the surroundings of the radio galaxy.  Jets and lobes therefore
provide a means for a mechanical coupling between accreting black hole
at the centre of the radio galaxy and its surroundings. In current
models of galaxy and black hole formation and evolution
\citep[e.g.][]{BBMea06,CSWea06,HRH06}, this AGN feedback is key to
regulating the growth of the black hole and establishing the locally
observed correlation between black-hole mass and large-scale galaxy
properties such as bulge mass and bulge luminosity
\citep{FM00,GBBea00,HaeringRix04}.

To calibrate and verify these galaxy formation models and to
understand the ICM thermodynamics, it is very important to determine
the magnitude of this mechanical energy input into a radio galaxy's
surroundings. Since the total energy of the radio lobes is dominated
by the lowest-energy electrons, low-frequency observations are
essential.  Yet, today, still very little is known about the shape of
the electron energy distribution at frequencies below 10-100 MHz.

\subsubsection{High-redshift galaxies and quasars}
\label{s:scidrv.extragal.hz}

One successful method to find high-redshift radio galaxies quasars in
the past was to search for ultra-steep-spectrum radio sources
\citep{RvOMea97}. Since observations at lower frequencies include
preferentially steeper-spectrum sources, a ULW survey might turn up a
large number of high-redshift radio sources.  Extrapolating (perhaps
optimistically) from source counts at 74\,MHz in the VLSS
(eq.~\ref{eq:Ncounts10MHz}), our example array from the same section
(300 elements, 100\,km maximum baseline, 10\,MHz observing frequency)
would thus be able to discover 3 Million sources down to the confusion
limit of 65\,mJy at $5\,\sigma$ significance over half the sky
(500\,000 per steradian), which it would be able to survey within
roughly two years.  For reference, \citet{RvOMea97} found 30 radio
galaxies at $z>2$ starting from a sample of $\approx700$ ultra-steep
spectrum sources.

Another, not yet well-explored method to identify active black holes
and jets in the early universe could be to look for young jets that
are still contained within the just forming galaxy. They are expected
to show up as compact radio sources with synchrotron self-absorption
turnover frequencies that are unusually low for their sizes
\citep{2004NewAR..48.1157F}. In the local universe these sources are
known as Compact-Steep-Spectrum (CSS) and GHz-Peaked-Spectrum (GPS)
sources \citep{ODea1998}, with turnover frequencies in the range
100-1000 MHz. At $z=10$ these would be shifted to 10-100 MHz and could
be readily identified by an ULW array.  Interestingly, the only
radio-loud quasar so far identified at $z>6$ may be a CSS source
\citep{FGPG08}.

The number-counts estimate above, of course, includes all kinds of
radio sources, so that the 500\,000 sources per steradian will also
include objects from this category and the ones described next.

\subsubsection{Fossil radio galaxies} 
\label{s:scidrv.extragal.oldRG}

Radio galaxies and other AGN are now believed to be normal galaxies in
a phase of more or less strong accretion onto their central black hole
\citep{CSWea06}. It is an important question how long this active
phase lasts, and on which timescale their activity recurs.

In the so-called ``double-double'' radio galaxies with two or even
sets of radio lobes \citep[e.g.][]{SBRea00}, the recurrence is
directly visible. From synchrotron ageing arguments, the age of the
outer set of lobes is inferred to be of the order of several tens of
millions of years, while the inner lobes are inferred to have been
launched after an interruption of several to several tens of millions
of years.  Thus, both the duration of an activity phase and the
recurrence timescale are likely within an order of magnitude of
$10^{7}$\,yr.

However, double-double sources are rather rare, perhaps implying that
special circumstances might be required for their formation. Other
methods are needed to determine the duration of a galaxy's active
phase in the general case. A very direct way of doing so is counting
how many ``fossil'' radio galaxies there are, i.e., those galaxies
whose activity phase has switched off recently.  However, the fossil
galaxies have been elusive so far, in part because the emitted power
of a radio galaxy drops as it grows in linear extent \emph{even} for
constant jet power \citep{BRW99}. This creates a ``youth-redshift
degeneracy'' in current flux-limited low-frequency radio surveys: only
fairly young radio galaxies (ages $\sim 10^5$--$10^7$\,yr) are
observable at high redshifts ($z\ga 0.8$), while at lower redshifts,
radio sources are observable in principle up to ages of
$10^8$--$10^9$\,yr \citep[Fig.~17]{BRW99}.  Because of this degeneracy
no strong statements can be made yet about radio source lifetimes
based on the study of low-frequency samples.

Moreover, the only known radio galaxy relicts have been identified as
\emph{cool holes} in the X-ray emission of the hot cluster gas
\citep[e.g.][]{McNamaraea01,FSTea06}, not via direct searches for
their radio emission (see also the following section on galaxy
clusters).  Present observations simply may have been done at
frequencies that are too high to detect fossil radio galaxies in
high-frequency all-sky surveys. This is likely to be the case also
with the much better sensitivities of the upgraded E-VLA and E-MERLIN
arrays. Lower-frequency observations are better suited for detecting
relict radio sources directly because they probe electrons with lower
energies, which are those with the longest radiative loss times. This
is because of two key properties of synchrotron radiation:
\begin{enumerate}
\item The frequency at which a relativistic electron of energy $E$ in
  a magnetic field of flux density $B$ emits synchrotron radiation
  scales as 
  \begin{displaymath}
    \nu_{\mathrm{c}} \propto E^2 B,
  \end{displaymath}
  i.e., lower-frequency emission is generated by lower-energy
  electrons.
\item The energy loss of a synchrotron-emitting electron scales as
  \begin{equation}
    - \frac{\de E}{\de t} \propto E^2 B^2, \label{eq:syndEdt}
  \end{equation}
  i.e., synchrotron cooling times are inversely proportional to
  energy.
\end{enumerate}
But it is not only the case that synchrotron cooling times increase
with decreasing observing frequency -- more strongly yet, there is a
\emph{maximum energy} that a synchrotron-cooling electron can have
after a given time has elapsed since its acceleration \emph{even if it
  had been given an infinite amount of energy initially}
\citep{vdLP69,Jes01}, because eq.~(\ref{eq:syndEdt}) integrates to
$1/E(t)-1/E(t_0) \propto t-t_0$.

Hence, observations at lower frequencies than currently accessible
allow fossil radio galaxies to be detected for a longer interval
between the switching-off time and the time of observation, with the
accessible radio-lobe lifetime increasing roughly inversely
proportional to the observing frequency.  Low-frequency radio
observations may even be the \emph{only} possibility for detecting
relict radio galaxies.  Even the continued non-detection of such
fossil galaxies would yield new physical insights, since it would put
the strongest constraints on how rapidly radio lobes have to dissipate
all of their energy after the central engine ceases to be active.

\subsubsection{Galaxy clusters}
\label{s:scidrv.extragal.clusters} 

Just like the gas in radio galaxies, the plasma that fills galaxy
clusters emits synchrotron radiation, as well as X-rays through
inverse-Compton scattering. The origin of the radiating relativistic
particles is unknown. Theories for their origin can be constrained by
low-frequency observations, analogous to those of radio galaxies. This
will yield important constraints on the formation history of the
clusters and of their magnetic fields. \citet{CBS06,CBVea08} have
shown that the steep spectra of cluster halos mean that observations
at lower frequencies are much more likely to detect the cluster halos,
specifically for low-mass and high-redshift objects.  Indeed,
\citet{BGCea08} report the detection of diffuse steep-spectrum
emission from the cluster Abell~521 at 240--610\,MHz that is not
detectable at 1.4\,GHz.  Again, we stress that even a fairly moderate
array of 300 dipoles and 100\,km maximum baseline operating at 10\,MHz
would be able to discover hundreds of thousands of sources in a matter
of a few months.

Independently of the question of the impact of AGN feedback on galaxy
evolution discussed in \S\ref{s:scidrv.extragal.edist} above, the
mechanical energy input into the intra-cluster medium (ICM) from giant
radio sources is an important quantity in studying the thermodynamics
of galaxy clusters, and in erudating how energy is transported through them in
order to establish the observed temperature and entropy profiles.  For
example, the detailed X-ray observations of the Centaurus and Perseus
clusters by \citet{FSTA05,FSTea06} show that bubbles blown by the
central radio source play an important role.  In the case of the
Perseus cluster, the X-ray images have revealed sets of ``ghost
bubbles'' without detectable high-frequency radio emission, which are
identified as relict sources as discussed in the previous section.
These ghost bubbles rise buoyantly in the cluster atmosphere and are
therefore part of the energy transport mechanism in clusters.  With
low-frequency observations, it may be possible to detect radio
emission from known bubbles, or new bubbles in clusters that have not
or cannot be observed to similar X-ray depth as Centaurus and
Perseus. In either case, low-frequency radio observations would yield
the most stringent constraints on the ages of such bubbles.

\subsection{Galactic/ISM surveys}
\label{s:scidrv.galaxy}

In this section we discuss how to infer the structure of the
interstellar medium and the origin of cosmic rays through
low-frequency observations. Beyond these, also the study of pulsars
may benefit from these because a confusion limit of 65\,mJy at 10\,MHz
with 1\arcmin\ resolution would be matched to the flux limit of LOFAR
for the $\nu^{-3}$ spectrum typical of pulsars. This implies that
pulsars could either be detected directly, or limits can be set on
their low-frequency cutoff.  However, the impact of temporal
broadening in the ISM (\S\ref{s:constr.gal.temporal}) on the
low-frequency observability of pulsars needs to be assessed in more
detail than we can achieve here.

\subsubsection{Structure of the ISM -- the Solar System's
  neighbourhood}
\label{s:scidrv.galaxy.ism}

\begin{figure*}
\begin{center}
\includegraphics[width=0.49\hsize]{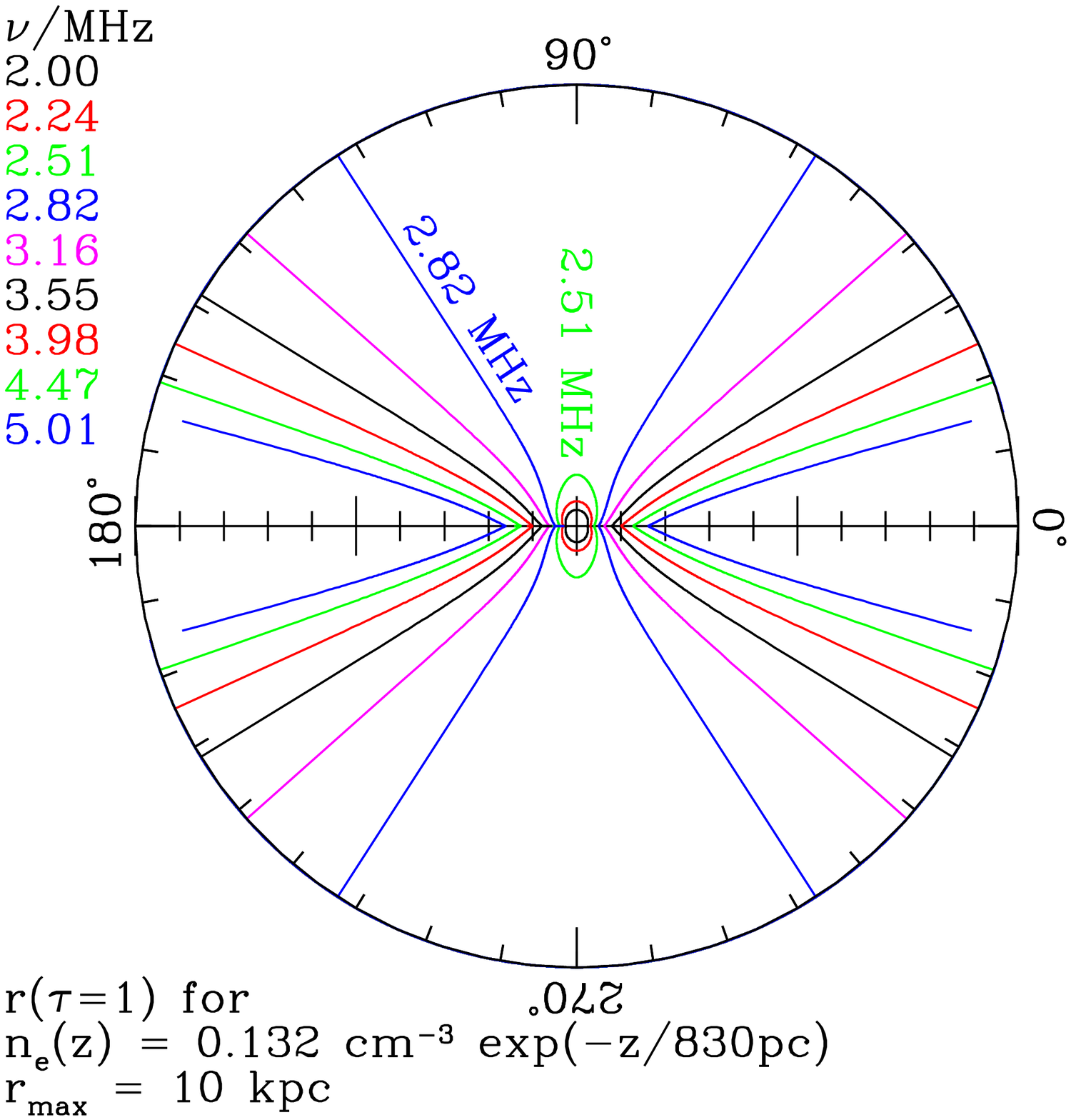}
\hfill
\includegraphics[width=0.49\hsize]{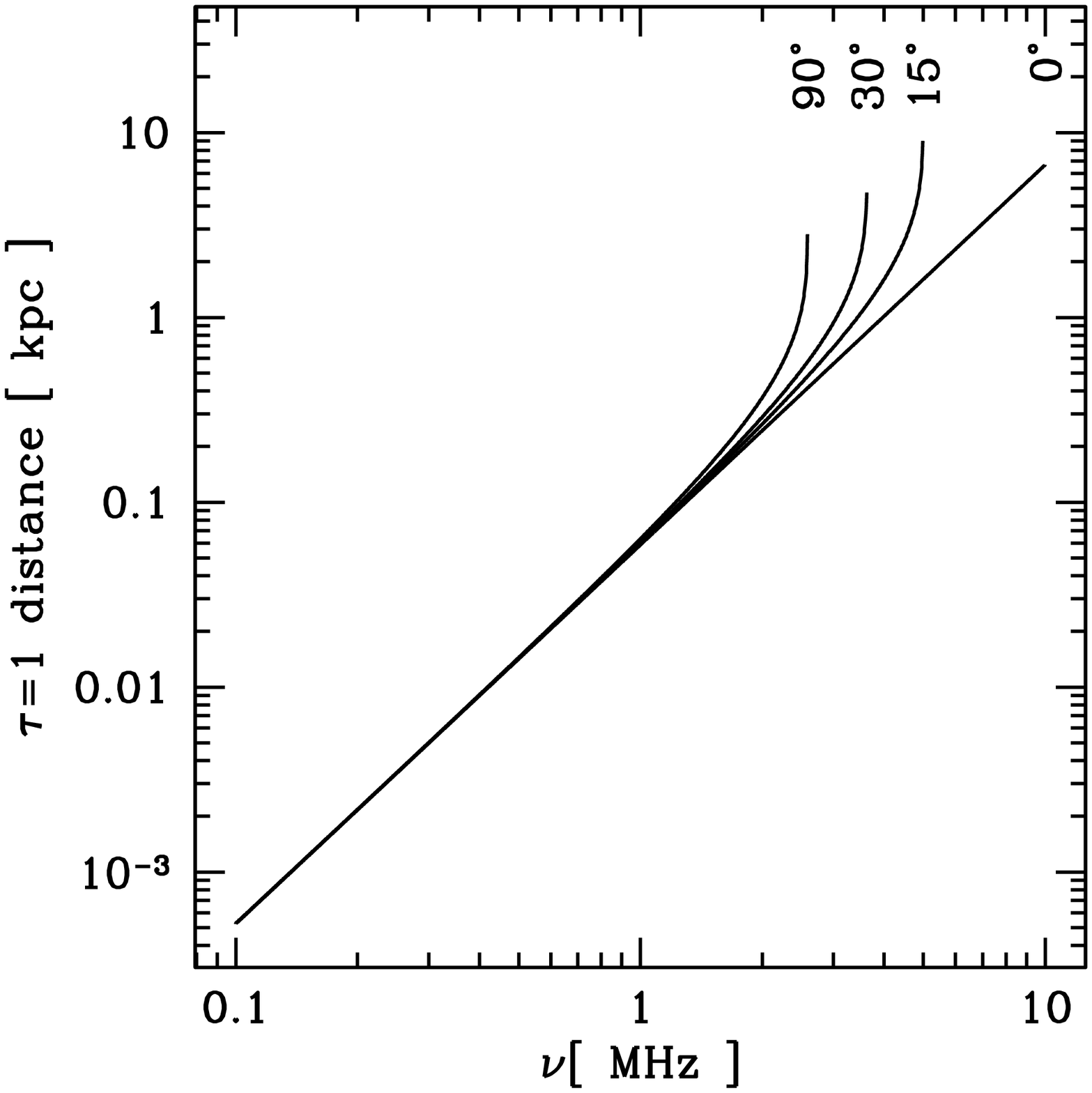}
\end{center}
\caption{\label{f:ISMpathlength}\textbf{Left:} Distance to $\tau=1$
  surface of the Warm Ionized Medium component of the ISM due to
  free-free absorption, shown as function of Galactic latitude for
  different frequencies. The figure assumes the exponential electron
  density profile shown and neglects any azimuthal structure of the
  ISM. The radius of the circle is 10\,kpc, and the galactic plane is
  along the horizontal. The density at the plane was chosen to match
  the transition between total absorption and unabsorbed sight lines
  in the polar directions to the observed turnover frequency of the
  ISM emission; for a more detailed approach, see
  \protect\citet{PW02}. The plotted profiles will be modified in
  practice by a higher electron density towards the Galactic centre,
  and by taking into account the detailed substructure of the
  ISM. \textbf{Right:} Distance to $\tau=1$ surface assuming the same
  azimuthally symmetric exponential electron density profile as in the
  left-hand panel, now as function of frequency for different angles
  above the midplane.  Around the turnover frequency of 2\,MHz, the
  sky is uniformly foggy; at lower frequencies, the galactic plane is
  seen in absorption against the brighter high-latitude emission,
  while above the cutoff, the ISM becomes less foggy, beginning at
  high frequencies and high galactic latitudes.  The limiting power
  law is given by $D(\tau=1) \approx 500\,\mathrm{pc}
  (\nu/\mathrm{MHz})^2$ (see
  eq.~\protect\ref{eq:visibility_ism}). Again, substructure will
  modify the $\tau=1$ distance along different lines of sight.}
\end{figure*}
The structure of the ISM in general, and of the supernova-inflated
``local bubble'' within which the solar system resides, has been inferred by considering
absorption lines in ultraviolet and X-ray spectra of stars and the ISM
\citep[e.g.]{Frisch98,Breitschwerdt01,OJSea05,HSK07} and the
distribution of pulsar dispersion measures in the sky \citep{CL02}.

Also low-frequency radio observations have been brought to bear on
this problem: \citet{PW02} have modeled the clumpiness of the free
electron distribution in the \emph{warm ionized medium} (WIM) phase of
the ISM by matching the expected turnover in the spectrum of the
Galactic synchrotron emission to that observed by the RAE-2 (see
\S\ref{s:intro}) and IMP-6 \citep{Brown73} satellites. Their method
models the radiative transfer through the WIM and requires only
knowledge (or reasonable assumptions) about the intrinsic, unabsorbed
shape of the underlying synchrotron emission spectrum in addition to
the observed brightness distribution. It is applicable over the
frequency range where the WIM is neither completely transparent nor
completely opaque, i.e., from 0.1--10\,MHz according to \citet{PW02}.

This modeling can be extended to infer not only the clumpiness of the
ISM, but also the full three-dimensional structure of the WIM.
Figure~\ref{f:ISMpathlength} shows the dependence of the ISM
``visibility'' (distance to the $\tau=1$ surface) as function of
Galactic latitude for different frequencies, and as function of
frequency for different wavelengths.  The limiting power-law for the
line-of-sight distance as function of frequency (right-hand panel in
Fig.~\ref{f:ISMpathlength}) is given by the inverse of the
free-free-optical depth (eq.~\ref{eq:kappa_ff}) for the electron
density and temperature in the midplane of the Galaxy; for the
parameters given in the figure and at wavelengths between 0.1 and
1\,MHz, we obtain
\begin{equation}
D(\tau=1) \approx 100\,\mathrm{pc}
\left(\frac{n_e}{0.132\,\mathrm{cm}^{-3}}\right)^{-2}
\left(\frac{T_e}{7000\,\mathrm{K}}\right)^{3/2}
\left(\frac{\nu}{1\,\mathrm{MHz}}\right)^2.
\label{eq:visibility_ism}
\end{equation}
By scanning through these frequencies and modelling the observed
emissivity as function of Galactic coordinates, the structure of the
WIM in the solar system's neighbourhood can be determined in 3D.  The
requirements on resolution and sensitivity are rather modest: taking
100\,pc as a typical ISM bubble size and 1\,kpc as a typical distance
leads to a degree-scale resolution requirement.  The sensitivity
requirement is comparatively modest also, since the ISM emission
itself sets the background temperature, so that a sensitivity of order
of one part in 1000 of the background level is sufficient, i.e., the
requirement is $\TRMS \approx 10^4$\,K near 1\,MHz.

Around the turnover frequency of 2\,MHz, the sky is uniformly foggy;
below it, the $\tau=1$ distance scales as given by
eq.~(\ref{eq:visibility_ism}). Above it, the ISM becomes transparent
at lower and lower angles above the midplane as the frequency is
increased, i.e., the turnover frequency decreases with the angle above
the midplane.  Around 2.5\,MHz, the entire Galaxy is transparent, while
at 0.3\,MHz, a frequency well above the putative lunar ionosphere, one
would see out only to a few tens of parsec, and see the Galactic plane
in absorption against the brighter WIM emission. At these low
frequencies, the visibility will therefore be set by the actual
substructure of the ISM (supernova shells and bubbles) rather than by
the global structure.

On the upside, the short distances out to which we can see at low
frequencies imply the effects of ISM angular and temporal broadening
are much less severe than for extragalactic observations at higher
frequencies.  For example, from the NE2001 model \citep{CL02}, the
pulse broadening time at 1\,MHz over a distance of 500\,pc in the
direction of the galactic pole is about 90\,s, instead of a few days
for extragalactic observations at the same frequency. Similarly, the
angular scatter broadening is 0.5\arcmin\ instead of
$\approx30\arcmin$, affording a linear resolution of 0.1\,pc at the
$\tau=1$ distance of 500\,pc (see Fig.~\ref{f:exoscatt} below for an
indication of how angular and temporal broadening scale with distance
and frequency).


\subsubsection{Origin of cosmic rays}
\label{s:scidrv.galaxy.crs}

Following the overview by \citet{Duric00}, this section describes an
approach for elucidating the origin of cosmic rays, i.e., relativistic
electrons and protons, with the help of a low-frequency radio array.

The essence of the approach suggested by \citeauthor{Duric00} is to
make use of the fact that Galactic HII regions are optically thick at
frequencies below about 30 MHz. Therefore, radio emission observed
towards such HII regions at low frequencies predominantly arises from
material along the line of sight to the HII region, i.e., the
synchrotron emission from cosmic-ray electrons. By observing the
synchrotron emissivity towards different HII regions at different
distances from the Earth, a 3D map of the electron density
distribution can be built up. To remove the degeneracy in the
synchrotron emissivity between electron density and magnetic field
strength, the radio data can be compared with gamma-ray
observations. 

Supernova remnants (SNRs) are known to accelerate high-energy
particles through the first-order Fermi shock acceleration
mechanism. However, the so-called ``injection problem'' has not been
solved, as the Fermi mechanism can boost particles which are already
mildly relativistic to very high energies, but it cannot accelerate
thermal particles to relativistic energies. Since the frequency of
synchrotron emission scales with the particle energy, low-frequency
radio observations can be used to trace the energy distribution of the
lowest-energy particles. This is necessary to constrain theories which
attempt to solve the injection problem.

The level of turbulence generated by SNRs in the interstellar medium
is also relevant to the generation and propagation of cosmic rays. By
observing the scatter broadening of extragalactic background sources
at various distances from a SNR, the level of turbulence can be probed
and mapped in 3D.

However, the resolution required to clearly resolve HII regions and
SNRs is of order 1\arcsec, and the required sensitivity is
0.1\,mJy/square arcsecond. Both are a challenge for a lunar array and
may be better done by a ground-based telescope at higher frequencies
and resolutions.

\subsection{Transient radio sources: suns and planets}
\label{s:scidrv.transients}

Transient sources are particularly interesting in astrophysics because
they often indicate very high-energy phenomena, e.g., Gamma-ray bursts
(GRBs), supernovae and other stellar explosions, outbursts of accreting
black holes, neutrons stars and white dwarfs, reconnection events in
the solar corona, or the impact of electrons from the solar wind on
planetary atmospheres. Other sources show transient activity simply
because of rotation, most notably radio pulsars. Finally, at
low frequencies the possibility for detecting coherent emission
processes increases due to the macroscopic wavelengths. To allow the
detection of transients, high-resolution high-sensitivity observations
have to be possible within the timescale of the transient phenomenon.

\subsubsection{Solar and planetary radio bursts}
\label{s:scidrv.transients.planets}

Both the Sun and planets produce radio bursts, the sun through
magnetic activity in its corona, and the planets and their moons
through interaction of charged particles in the solar wind with the
planet's or moon's magnetic fields, as is the case for exoplanets (see
section \ref{s:scidrv.transients.extraplanets}). In the case of the
Sun, the most violent coronal events give rise to coronal mass
ejections (CMEs), disruptions in the solar wind in which large numbers
of charged particles (i.e., cosmic rays) are accelerated. The magnetic
activity and the charged particles can have dramatic effects on the
Earth and Moon, posing a danger to astronauts and electronic equipment
outside the Earth's atmosphere, or disrupting radio
communications. These effects have been called ``magnetic storms'',
and efforts to predict the strength and timing of CMEs impacting on
the Earth-Moon system are known as ``space weather prediction''. The
CMEs themselves are traceable via their low-frequency radio emission
as they propagate towards the Earth with an array capable of
performing imaging at high time resolution.  Since the emission
frequency is dependent on density and thus radius, observations down
to 100~kHz would trace emission out to about half an AU \citep[using
the density model of][e.g.]{MJMcD99}.

While the Sun is the strongest source of radio emission at high
frequencies (at 100~MHz and above), bursts from the magnetized
planets, in particular Jupiter, can reach a strength comparable to
that of the solar bursts or more, i.e., a few MJy \citep{Zarka07}. The
emission region of Jupiter is very small and not resolved on
lunar-like baselines \citep[see][]{NZKea07}. Hence, an imaging
application for Jupiter radio emission is not a scientific driver for
a lunar telescope. However, since the emission is strong enough, a
single broad-band receiver covering frequencies $\la30$ MHz for
spectral measurements with high frequency and temporal resolution of
solar and Jupiter bursts could be envisaged for prototyping and
background measurements. For angular discrimination between different
burst sources, a small interferometer with comparatively modest
baselines (beginning at 500\,m for the highest frequency of 30\,MHz)
would be sufficient. In the long term, also interferometry on a
Moon-Earth baseline could be considered to constrain Jupiter bursts in
new ways.

\subsubsection{Extrasolar planets}
\label{s:scidrv.transients.extraplanets}

While Jupiter's emission is an additional foreground and interference
signal that likely needs to be avoided for high-sensitivity
low-frequency observations, it implies that low-frequency radio
observations offer the most favourable achievable intensity contrast
between an extrasolar planet and its sun, and therefore open up the
possibility of a direct detection of extrasolar planets and their
magnetosphere.

\citet{Zarka07} has reviewed the radio emission from extrasolar
planets and the results of searches to date.  Magnetized planets emit
bursts of low-frequency radio waves through the cyclotron maser
instability (CMI), in which electrons from the stellar wind interact
with the planet's magnetic field.  The CMI is 100\% circularly
polarized, while solar/stellar bursts are usually unpolarized,
offering a mechanism of distinguishing between stellar and planetary
radiation.  Without a polarization-sensitive array, planetary and
solar bursts would be distinguishable by temporal structures, either
within bursts or from modulations on the timescale of the planetary
orbit.  \citet{LFDea04} give 15\,minutes as typical burst duration,
while \citet{Zarka07} shows an example of a burst with an overall
duration of several hours, with frequency structure down to the
millisecond level (in his nomenclature, only the features on
sub-second timescales are called ``bursts'').

Since the CMI is only active in the presence of strong magnetic
fields, either in the planet or the stellar wind, this mechanism is
most suitable for the detection of magnetized Jupiter-like planets.
Earth-like planets will only give rise to radio bursts if they are
embedded in a stellar wind which is itself strongly magnetized.  Even
if this is a rare situation, the radio emission might be the most
readily detectable signal of such planets.

A crucial parameter is of course the rate at which planetary bursts
occur, and hence the survey time that needs to be invested in order to
obtain a statistically relevant sample of burst observations.
However, very little is known about the likely recurrence rate, which
depends on the unknown way in which CMI bursts are triggered. One
possibility is that they have ``light house beams'' similar to
pulsars, with intrinsically continuous emission that is confined to a
solid angle and bursts being observed when the emission beam sweeps
over the telescope \citep[e.g.,][]{ZCK04}.  Another possibility is
that CMIs are triggered or enhanced by coronal mass ejections
\citep[CMEs; see][who give an example of a solar CME enhancing Jovian
  radio emission]{GKHea02}. In this case, the relevant timescale is
the CME recurrence timescale modified by the probability that a CME in
fact interacts with the planet.  Based on such considerations,
\citet{KRLea07} expect CME rates in excess of the solar value of about
6~CMEs/day, in particular for M~dwarfs which are intrinsically more
active than solar-type stars. \citet{GPKea07} add that CMEs can lead
to an effectively continuous enhancement of planetary radio emission,
in particular in young stars with their frequent CMEs.

In spite of the current uncertainties about the rate of exoplanetary
low-frequency bursts, the possibility of a direct detection of
exoplanets via such bursts is an exciting prospect. Being closer to
their star than Jupiter is to the Sun, hot Jupiters should show much
more intense radio emission than Jupiter itself, at least of order
$10^5$ times more \citep{LFDea04,GMMea05,Zarka07}.  In particular,
\citet{LFDea04} have computed the expected radio emission for the
extrasolar planets known at the time (mostly hot Jupiters) and found
that they should have bursts of flux densities close to 1~mJy at
frequencies of several tens of MHz up to a few GHz, while a few
lower-mass planets should emit at frequencies around 0.1-1\,MHz and
reach tens to hundreds of mJy.  Recently, \citet{GZS07} have compared
predictions for three separate emission processes for currently known
exoplanets, and find lower peak frequencies and average emitted powers
than \citet{LFDea04}.

\begin{figure*}
  \begin{center}
\includegraphics[width=0.55\textwidth]{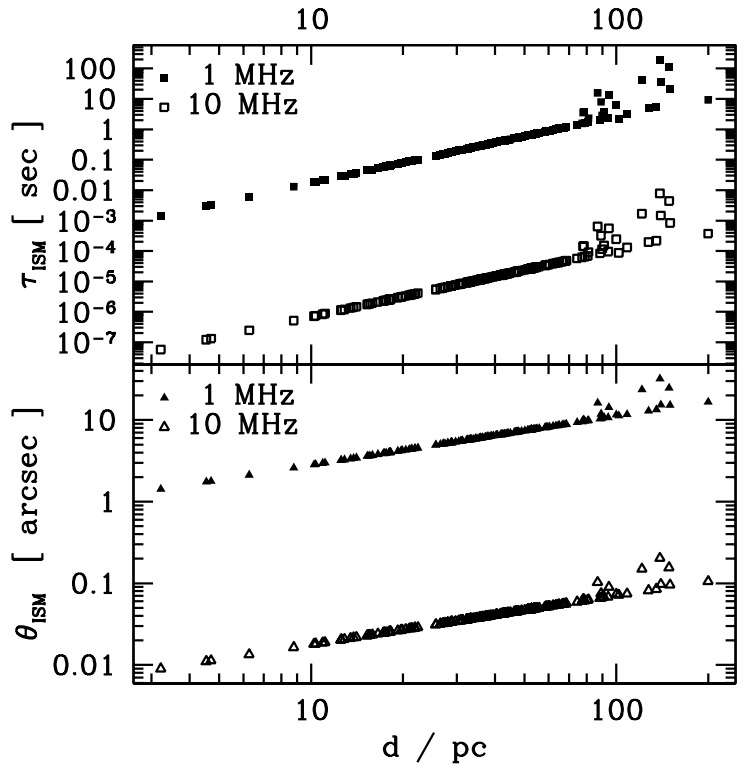}
\end{center}
  \caption{\label{f:exoscatt}Pulse broadening times (top panel) and
    angular broadening scales (bottom panel) caused by ISM turbulence
    towards known extrasolar planets, as calculated from the NE2001
    model for the distribution of free electrons in the Galaxy
    \protect\citep{CL02}. The temporal and angular broadening is shown
    for the two different frequencies as indicated; the strength of
    the former scales as $\nu^{-4.4}$, that of the latter as
    $\nu^{-2.2}$.  The angular broadening is always negligible
    compared to the likely resolution of a lunar array (compare
    Fig.~\ref{f:resBLscatt}). The temporal broadening at 1\,MHz is
    short compared to the likely exoplanetary radio burst duration of
    $\approx$\,15 minutes for all but the most distant planets behind
    those ISM regions with the strongest turbulence.}
\end{figure*}
Referring to Fig.~\ref{f:sensNant}, the detection of a 1\,mJy burst at
$5\sigma$ significance over a typical duration of 15 minutes
\citep{LFDea04} at a frequency of 30\,MHz and at 25\% bandwidth
requires of order $10^5$ dipoles, while detecting a 10-mJy burst of
equal duration at 3\,MHz (and again with 25\% bandwidth) still
requires about $10^4$ dipoles; at fixed frequency, the required number
of dipoles is inversely proportional to the required
sensitivity. Source variability and their Galactic nature would help
to overcome the confusion limit. \label{p:planetdipoles} The
requirement for the number of dipoles is driven by the need to detect
a burst within its duration. Furthermore, there is no need to map the
spatial structure of a burst, i.e. the sensitivity requirement is on
flux, not surface brightness. Hence, the dipoles can be distributed
over baselines that yield arcminute-scale resolution, which is
sufficient for establishing positional correspondence to observations
at other wavelengths.

Before investing in the construction of such an array, it is important
to verify that the temporal pulse broadening due to ISM turbulence
(\S\ref{s:constr.gal.temporal}) does not smear out an exoplanetary
burst with a duration of $\approx$15 minutes beyond recognition.  We
used the Fortran implementation of the NE2001 model for the
distribution of free electrons in the Galaxy by \citet{CL02} to
determine the scattering regime, pulse-broadening times, and angular
broadening for 176 exoplanets\footnote{Taken from the list at
  \url{http://exoplanets.org}}. The planets have a mean distance of
45\,pc, and sample a large range of Galactic longitudes and latitudes
(though their distributions naturally follows that of Galactic stars,
so that they are concentrated towards the Galactic plane).

Figure~\ref{f:exoscatt} shows the pulse-broadening times and angular
broadening scales for the exoplanets as a function of
distance. Although all exoplanets are in the strong scattering regime
at frequencies below 30\,MHz, the angular broadening even at 1\,MHz
reaches at most a few 10s of arcseconds, which is negligible compared
to the likely resolution of a lunar array (compare
Fig.\,\ref{f:resBLscatt}).  The pulse broadening times at 10\,MHz are
nearly exclusively well below 1\,ms, and in all cases less than
10\,ms.  At 1\,MHz, the pulse broadening is a factor of $10^{4.4}$
larger (eq.~\ref{eq:tau_ISM}), but still below 1\,sec and hence
negligible for all planets up to 70\,pc distance. Fewer than 1/6 of
the known planets are at a greater distance than 70\,pc, and only 5 of
the 113 planets considered by \citet{LFDea04} are predicted to emit at
frequencies below 1\,MHz (\citealt{GZS07} do not report any
predictions below 1\,MHz because they considered only Earth-based
radio observatories).  Moreover, as noted above, there are
indications \citetext{P.~Zarka, \emph{priv.\,comm.}} that the pulse
broadening may increase less rapidly with wavelength than expected
from eq.~(\ref{eq:tau_ISM}).  But even if that equation holds,
Fig.~\ref{f:exoscatt} shows that angular and pulse broadening by the
ISM impose no significant constraints on the observability of
exoplanetary radio bursts.

\subsection{Ultra-High Energy particle detection}
\label{s:scidrv.uhe}

\subsubsection{Hadronic Cosmic Rays}
\label{s:scidrv.uhe.cosmics}

\begin{figure*}
\begin{center}
\includegraphics[width=0.55\textwidth]{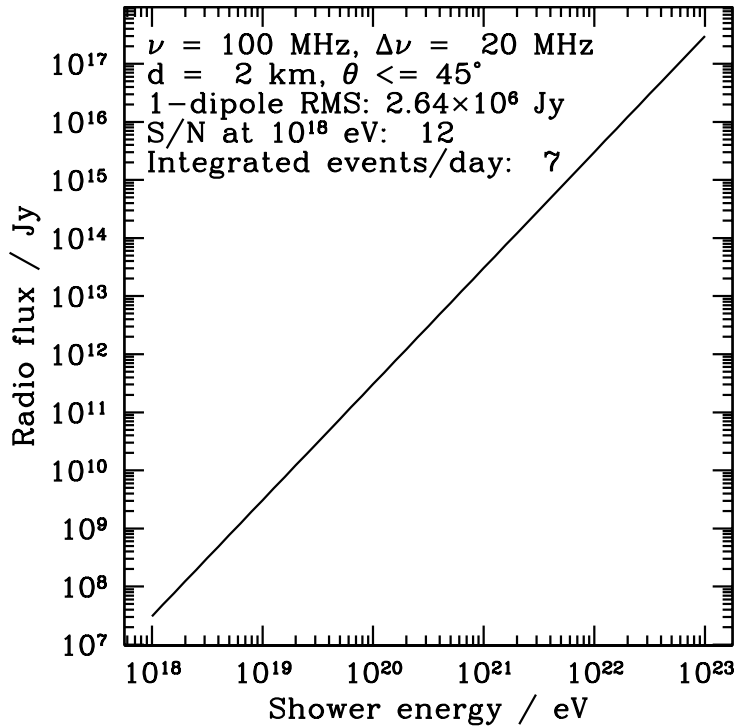}
\end{center}
\caption{\label{f:CRfluxdens}Radio flux from cosmic-ray or neutrino
  events interacting with the Moon, as observed at 100~MHz with a
  single dipole located on the Moon. The flux and event rate were
  computed using the formulae given by \protect\citet{SBBea06}.}
\end{figure*}

Ultra-high energy cosmic rays (UHECRs) are a rare population of
elementary particles, likely composed of protons, atomic nuclei, and
potentially photons or neutrinos \citep{2000RvMP...72..689N}.  Their
energies can reach much higher values than those currently achievable
in man-made particle accelerators.  The total flux of cosmic rays is
about 1 particle m$^{-2}$ sterad$^{-1}$ year$^{-1}$ above a fiducial
particle energy of 10$^{16}$ eV. The cumulative flux
spectrum\footnote{i.e., the total number of particles above that
  energy} drops roughly as $E^{-2}$ with higher energy thresholds,
flattening to $E^{-1.75}$ below the so called ``knee'' in the cosmic
ray spectrum at $2\times10^{15}$ eV \citep{2003APh....19..193H}. Above
the knee, the cosmic ray flux is roughly given by
\begin{eqnarray}\label{crflux}
F_{\rm CR}(E>E_{\rm c}) &=&30\, {\rm~particle}\, {\rm km}^{-2}\, {\rm
sterad}^{-1}\, {\rm day}^{-1} \; \left({E_c\over10^{17}{\rm
eV}}\right)^{-2}\\
& \mathrm{at} & E_{\rm c}>2\cdot10^{15}{\rm eV}.\nonumber
\end{eqnarray}
At the highest energies, the particle spectrum extends up to and
perhaps beyond $10^{20}$\,eV. The currently best information on the
spectrum is available from the Pierre Auger Observatory
\citep{2008arXiv0806.4302T}, showing the well-known ``ankle'' around
$4\times10^{18}$ eV and a steepening beyond $4\times10^{19}$ eV. The latter
is likely due to the Greisen-Zatsepin-Kuzmin (GZK) cutoff, a
suppression of extragalactic UHECRs due to proton-gamma interactions
with the cosmic microwave background. Origin and nature of UHECRs is
still largely unknown, but a link with nearby accreting supermassive
black holes has been made \citep{2007Sci...318..938T}. A major
obstacle to understanding these most energetic particles in the
universe is their small flux of less than 1 particle per square
kilometer per century, requiring large detectors. Here the moon, with
its large surface exposed to empty space, can play a significant role
as a particle detector target.

UHECRs can be detected through their intense particle cascades which
they initiate when encountering a target. For example, if UHECRs hit
the lunar surface a particle shower will pass through the lunar
regolith with a speed greatly exceeding the local speed of light (the
refractive index of the lunar regolith is $n=1.73$, see
\citealt{2006PhRvD..74b3007A} and references therein), leading to
intense Cherenkov light \citep{ZHS92,DZ89,HEOS96,GHLea04}. Due to the
finite length of the shower of a few meters, the emission must be
coherent at long radio wavelengths. The shower will then show up as a
coherent ultra-short radio burst on a time scale of $\Delta t=1/\Delta
\nu$, which is tens of nanoseconds up to microseconds in the
low-frequency regime ($\nu=1-100$ MHz).

Hence, cosmic rays will in principle create an impulsive noise
background on top of the steady Galactic noise, but the emission will
in turn also allow one to them in more detail. \citet{FG03} and
\citet{SBBea06} have shown that the radio emission from such lunar
showers is detectable at the very highest energies by a LOFAR-like
telescope even from Earth. Especially for cosmic rays and neutrinos at
energies $>10^{21}$\,eV, well beyond the GZK cut-off, Earth-based
low-frequency observations of lunar radio bursts generated by
interactions within the moon offer a unique possibility to study these
elusive particles.

However, interpretation of these results in the end depends on
subsurface properties of the moon and shower-development at these high
energies (e.g., due to the Landau-Pomeranchuk-Migdal effect), making
in-situ measurements invaluable. This is impossible on the Earth since
accelerators cannot produce particles at such high energies and the
Earth's atmosphere prevents UHECRs from directly hitting the surface.

Moreover, placing low-frequency antennas on the moon would have
several interesting advantages compared to Earth-bound radio
experiments: First, being 400,000 times closer to the location of the
shower (assuming a fiducial shower distance of 1 km on the moon) than
the Earth, would, roughly speaking, increase the sensitivity to cosmic
rays and neutrinos by a factor $400,000^2$ and allow correspondingly
lower-energy particles to be detected.  Secondly, lunar radio pulses
observed from the ground get dispersed by free electrons in the Earth
ionosphere which can be avoided with in-situ measurements. Also, the
expected radio-quietness at the lunar surface would greatly aid in
solidly detecting these events.

Using the equations given by \citet{SBBea06}, we find that a single
dipole antenna operating at 100 MHz with 20 MHz bandwidth should be
able to detect of order 7 cosmic-ray events per day at $10^{18}$\,eV
and above at flux densities of 100 MJy, corresponding to
signal-to-noise levels of 10 or more, assuming events at a distance of
or less 1\,km, arriving from a cone of half-opening angle 45\deg\ 
(Fig.~\ref{f:CRfluxdens}).

One advantage of placing the radio receivers directly on the moon
instead of on the Earth is that high-energy events, which are rarer,
can be detected over greater distances from the location of the
antennas.  Therefore, a larger fraction of high-energy events is
observed, and the differential energy distribution of the
\emph{observed} events scales as $1/E$, even though the
\emph{intrinsic} distribution scales as $1/E^3$; the difference arises
because more energetic events produce quadratically greater fluxes,
which are detectable over linearly greater distances, increasing the
collecting area quadratically with energy. Thus, a lunar array finds a
larger fraction of the rarer high-energy events than an Earth-based
array.  

The total rate of detected events increases logarithmically with
energy, up to a maximum given by the largest distance out to which an
event's radio pulse is still visible to the antenna(s).  The total
event count will still be heavily dominated by low-energy events, so
that the total number of detected events above some energy is
approximately given by the number of events at that energy which occur
within the radius for which the signal-to-noise ratio $S/N$ meets the
detection threshold.

At least three antennas are necessary to determine the location of a
cosmic-ray event by triangulation and hence the original energy of the
particle.  An array of three elements will achieve an S/N of about 2.5
times that in Fig.~\ref{f:CRfluxdens}, since the S/N scales as
$\sqrt{N (N-1)}$, where $N$ is the number of antennas. The total event
rate above some fixed $S/N$ scales as $\ln(N[N-1])$, because lower
minimum energies $E_{\mathrm{min}} \propto 1/\sqrt{N(N-1)}$ become
accessible and the differential event rate scales as
$1/E_{\mathrm{min}}$.

If the observing frequency is lowered from 100\,MHz to 10 MHz, the
flux density per event will be much lower, while the sky background is
slightly higher, and, of course, the available bandwidth is
smaller. All three effects mean that the S/N for a fixed event energy
is smaller at lower observing frequencies.  On the other hand, the
emission is expected to be more isotropic at lower frequencies; while
only events above $10^{20}$\,eV would be observable at 10 MHz,
\emph{all} such events are detectable out to a radius of 5\,km by a
single dipole, with a rate of one detectable event about every 2
years.  This means that observing cosmic rays is feasible in the ULW
region, and an astronomical array of $\approx 100$ elements with 5\,km
spacing could detect of order 30 events/year at $E\ga
10^{20}$\,eV. This is in principle competitive with the largest cosmic
ray detectors today.

\subsubsection{Neutrinos}
\label{s:scidrv.uhe.nus}

The arguments used for deriving the radio flux generated by hadronic
cosmic-ray events also apply to neutrino interactions, so that the
basic array parameters are identical for the detection of both
neutrinos and more massive cosmic rays. The basic difference to the
hadronic component of UHECRs is that neutrinos will penetrate deeper
into the moon before they interact. The mean free path is 130 km
$\left(E_{\rm cr}/10^{20}{\rm eV}\right)^{-1/3}$ and hence the radio
pulse would come from {\it deep below} the surface. Here, lower radio
frequencies have the additional advantage of penetrating deeper into
the moon crust, thereby enlarging the usable detector
volume. \citet{SBBea06} quote an attenuation length of
\begin{equation}
l=9\,{\rm m}\;(\nu/{\rm GHz})^{-1}.
\end{equation}
If this scaling continues all the way to 10 MHz, then a single dipole
can see a volume of $\sim1.5 {\rm km}^3$ of detector material for
neutrino interactions. Hence, a simple radio array could in principle
easily cover several tens of cubic kilometers. This would be more
difficult at these wavelengths on Earth, due the global short-wave
radio interference. However, neutrino searches have already been
proposed and performed over Antarctica at higher frequencies
\citep{BBBea06,Halzen07,Spiering07}. Therefore, the competitiveness of
a lunar array needs to be evaluated more thoroughly. It may also be
that the difference in geometry of the moon and the density of the
lunar crust leaves an interesting parameter space open.

These experiments could run in parallel with an imaging array. They
would require an additional signal path with a pulse detection
algorithm and a raw time series buffer of several microseconds to
milliseconds per antennas. The trigger event time has to be at the
tens of nanosecond level. Unlike for most other applications, the
antennas would require a large acceptance in the horizontal planes to
detect ``grazing-incidence'' pulses and toward the ground. For simple
dipoles without ground plane an up-down ambiguity remains; however,
bandwidth-limited pulses produced in the near field above the lunar
surface appear exceedingly unlikely, so that this up-down ambiguity
should not be a concern. The distinction between neutrinos and
protons/nuclei in the cosmic radiation would be made based on the
geometry: only neutrinos would come from well below the array.

As a side-effect of a lunar neutrino detector, one could observe the
Earth as a neutrino source. The emission is due to the production of
atmospheric neutrinos which are generated in particle showers
initiated by cosmic rays hitting the Earth's atmosphere\footnote{It is
  amusing to note that the level of neutrino emission from a rocky
  planet like Earth is in fact a signal for the presence of an
  atmosphere. Showers in a solid medium that is not shielded by an
  atmosphere produce far fewer neutrinos due to the much shorter
  shower length. Unfortunately the level of atmospheric neutrino
  emission is far too low to be used in the study of
  exoplanets.}. Since the terrestrial neutrino flux is well known, one
could wonder about using the Earth-Moon baseline for neutrino
oscillation measurements.

\subsection{Meteoritic Impacts}
\label{s:scidrv.meteo}
Finally, we point out the possibility that impulsive radio noise
measurements might also be a potential tool to investigate meteoritic
impacts on the moon. Such impacts seem to produce optical flashes
which are visible even from Earth
\citep{2000Natur.405..921O,CDPea03,Cudnik07}. It is not inconceivable
to think that such impacts are also accompanied by low-frequency radio
emission. This may be in analogy to strong electromagnetic pulses
(EMPs), known from nuclear explosions, which have already been used to
explain radio emission from cosmic rays in the Earth's atmosphere
\citep{Colgate67}. One could also imagine an interaction between the
explosion and local electrostatic or magnetic field, however, all of
this remains speculative at the moment.

At least, radio emission from impacts has already been used by radio
experiments in free space, where radio antennas exhibit radio pulses
due to high-speed impacts of dust grains. For example, based on early
findings of Voyager II, a radio antenna on the Cassini spacecraft was
used to measure dust particles from Saturn's rings hitting the entire
spacecraft \citep{Kurth06}. The idea here is that the energy of the
impact creates an expanding ionized plasma with an ambipolar electric
field, which in turns creates a strong voltage pulse on the antenna or
spacecraft. The time scale of the pulse is on the ms scale and hence
should be well discernible from CR impacts. Moreover, the frequency
spectrum of the radio pulse seems related to the size of the impacting
particle. The radio signals might therefore reveal the size and mass
distribution of the particles. However, this effect typically is
associated with electrostatic noise generated at or near the antenna
itself. In how far this effect can applied to a lunar antenna remains
unclear at present.

\section{Summary \& Discussion}
\label{s:summary}

In this paper we have summarized some of the main science goals,
requirements, and limitations for astronomy at very low radio
astronomy frequencies below 100 MHz, i.e., ultra-long wavelengths
(ULW). In this frequency range the moon, in particular its far side,
offers by far the best (albeit not cheapest) site for conducting
observations. In several cases a lunar telescope even offers unique
insights.

The main limitation for ground-based low-frequency telescopes is
degradation of the image quality due to the Earth's ionosphere in the
10-100 MHz frequency range.  At very low frequencies ($<10$--30 MHz),
the ionosphere is even blocking our view of the universe
completely. Space and in particular lunar radio arrays are therefore
back in the focus. Here we have discussed a number of science areas
where this frequency range may be of special interest: cosmology with
primordial HI at $z=10$--100, extragalactic surveys of radio galaxies,
clusters and galaxies, astroparticle physics of cosmic rays and
neutrinos hitting the moon, transient flares from stars and
exoplanets, the 3D structure of the local interstellar medium, and
solar system objects.

\subsection{Science goals}

\begin{sidewaystable*}
\caption{Overview of science cases and requirements.\label{t:overview}}
\begin{tabular}{llrrrrrr}
\hline 
\multicolumn{1}{c}{Topic} 
& \multicolumn{1}{c}{Our \S} & \multicolumn{6}{c}{\textit{Requirements}} \\
& & \multicolumn{1}{c}{Frequency} &
\multicolumn{1}{c}{Resolution $\vartheta$} &
\multicolumn{1}{c}{Baselines} 
& \multicolumn{1}{c}{Expected signal\footnote{Surface
    brightnesses given in Kelvin, fluxes in Jansky}} 
& \multicolumn{1}{c}{N(Antennas)} &
\multicolumn{1}{c}{$t_{\mathrm{exp}}$ (5$\,\sigma$)} \\
& & \multicolumn{1}{c}{MHz} & & \multicolumn{1}{c}{km} & & &  \\
\hline
\textbf{Cosmology} & \textbf{\ref{s:scidrv.cosmo}}  &&&&&& \\
Global Epoch of Reionization\footnote{Here we use $10<z<20$ as fiducial
  redshift limits for the Epoch of Reionization (EoR), and $30<z<50$ 
  for the Dark Ages..} & \ref{s:scidrv.cosmo.gEoR} & 50--150
& $2\pi$\,sr & 0 & 5--50\,mK & 1 & 2\,h--20\,d\\
Global dark-ages signal & \ref{s:scidrv.cosmo.globaldark} & 30--45
& $2\pi$\,sr & 0 & 0--40\,mK & 1 or more 
& 0.3--30\,yr\\
EoR tomography & \ref{s:scidrv.cosmo.tomo} & 50--150  & $\approx
10\arcmin$--1\arcmin & 0.7--20 & $\approx 1\,$mK & $10^{5}$--$10^{7}$ 
\protect\footnote{This requirement is driven by the demanding the necessary
  surface brightness sensitivity over timescales compatible with the
  human lifetime, i.e., of order of years. Note the upper limit N(Antennas) $<
  N_{\mathrm{max}} = 4.7\times 10^{7}
  \left(\vartheta/1\arcmin\right)^{-2}$ from the shadowing limit
  (eq.~\ref{eq:Nmax}).  The array parameters for 21-cm tomography and
  power-spectrum observations scale as $N_{\mathrm{Ant}}\sqrt{t_{\mathrm{exp}}}\, S^{-1}\, \vartheta^{2}
  (1+z)^3 = \mathit{const}$, where $S$ is the desired sensitivity. \label{fn:sens}}
& yr\\
EoR 21-cm power spectrum & \ref{s:scidrv.cosmo.powspec} & 50--150 &
$\approx 2\deg$--$2\arcmin$ & 0.05--10 & $\approx 0.3$--5\,mK & $10^3$--$10^{5}\;\mpfootnotemark[3]$ 
& yr\\
Dark-ages 21-cm power spectrum & \ref{s:scidrv.cosmo.powspec} & 30--45 &
$\approx 20\arcmin$--$2\arcmin$ & 1--20 & $\approx 0.03$--1\,mK
& $10^{4}$--$10^{8}\;\mpfootnotemark[3]$ & yr\\
\hline
\textbf{Extragalactic surveys} & \textbf{\ref{s:scidrv.extragal}} &
10 & 1\arcmin & 0.1--100 & $\approx 65\,$mJy\footnote{Confusion limit
  from eq.~(\ref{eq:conflim}); about $3\times 10^6$ sources are expected to
  this limit in a  half-sky survey.} & 300 & 2\,yr\\
\hline
\textbf{Galactic surveys} & \textbf{\ref{s:scidrv.galaxy}} & & & & & & \\
Solar System neighbourhood & \ref{s:scidrv.galaxy.ism} & 0.1--10 &
degrees & 0.3--30 & 10000\,K & 10--100 & yr \\
Origin of cosmic rays & \ref{s:scidrv.galaxy.crs} & 0.1--30 &
1\arcsec & (3--30)$\times10^3$\footnote{Moon diameter: 3,476\,km} &
155,000\,K & 100,000 & 
 100\,d\\
\hline
\textbf{Transients} & \textbf{\ref{s:scidrv.transients}} & & & & & & \\
Solar and planetary bursts & \ref{s:scidrv.transients.planets} &
0.1--30 & degrees & 0.5--200 & MJy & 1--100 & minutes--hours\\
Extrasolar planets & \ref{s:scidrv.transients.extraplanets} &
0.5--30 & $\la 1\arcmin$  & $\ga 35$--1000 & 1--10\,mJy & $10^4$--$10^5$\footnote{The number of antennas is driven by requiring a
  5-$\sigma$ burst detection within the burst typical burst duration
  of 15 minutes, with $t_{\mathrm{exp}}\, N_{\mathrm{Ant}} = \mathit{const}$.} &
15\,min\\ 
\hline
\textbf{Ultra-High Energy particles} & \textbf{\ref{s:scidrv.uhe}}
& 10--100 &
N/A\footnote{Near-field triangulation of event sites helps, but is not
  strictly required.} & 0--5 & 100\,MJy &
1--100 & N/A (bursts)\footnote{Expected event rate for hadronic cosmic
  rays: $\sim7$/day at
  $E>10^{18}$\,eV for 1 dipole at 100\,MHz; $\sim30$/year at
  $E>10^{20}$\,eV for 100 dipoles with 5\,km spacing at 10\,MHz}  \\
\textbf{Meteoritic Impacts} & \textbf{\ref{s:scidrv.meteo}} & & & & & & \\
\hline
\end{tabular}
\end{sidewaystable*}
Table~\ref{t:overview} summarizes the various science goals and
requirements.  Clearly of greatest interest is the question of the
dark ages and HI observations at the highest redshifts. This would
contain a huge amount of original information about the very early
universe unaltered by the complicated astrophysical processes
associated with cosmic reionization.

For example, a determination of the exact boundaries of the Epoch of
Reionization through its global signal could in principle be achieved
with a single antenna in a dark crater near one of the lunar
poles. This would provide almost perfect shielding from any radio
interference and temperature (hence gain) variations -- ideal
conditions for such measurements. However, this experiment is also
being tried from the ground under somewhat less than ideal conditions
and the outcome has to be awaited (e.g., see the loose upper limit
just obtained by \citealt{BRH08} with the EDGES experiment).

The next step up is the measurement of the HI power spectrum above
$z>10$ and into the dark ages. Since the HI gas will trace directly
the dark matter distribution, this is an excellent way to measure the
primordial density fluctuations which led to the formation of large
scale structures in the universe. 

Measuring the baryonic acoustic oscillations (BAOs) also during the
dark ages has several advantages, the most important one being the
lack of non-linear processes and biases. Non-linear effects, e.g., the
rapid growth of ionization bubbles from the first stars, affect HI
measurements during the epoch of reionization, while unknown star
formation efficiencies affect measurements at lower redshifts using
galaxies. Hence, measurements of BAOs in the dark ages could tackle,
for example, the issue of non-Gaussianity in the density fluctuations
\citep{CoorayLiMelchiorri08}. This is a quantity directly revealing
information about the inflationary phase of the universe. Measuring
the BAOs at $z=50$ on a scale of tens of arcminutes requires of order
$10^{5}$ antennas, because the amplitude of BAOs is at the scale of
10s of microKelvin.  If one tries to push the resolution further down
to arcminute scales in order to discriminate between cosmological
models as discussed by \citet{LZ04}, one again requires at least of
order $10^5$ antennas at $z\ge20$. This is a somewhat futuristic
experiment for a space array. If this is realized with 5\,m crossed
dipoles one requires about 1000 km of wires, which corresponds to of
order 1 ton of metal for a light weight wire of one gram per
meter. While this is not the limiting factor, the power consumption
and the electronics might be. Such broad-band low-frequency receivers
currently require a power of the order of several Watts, bringing the
total energy consumption rate up to several 100\,kW plus a correlator
-- hence this would require some serious but not completely outrageous
lunar infrastructure to be available.

The main obstacle for achieving the same science goal from the ground
is the ionospheric seeing -- not because of the resolution, but
because of calibration problems. The detection of the HI signal
against the enormous sky background requires a perfect removal of all
strong confusing sources and their side lobes. Since low-frequency
observations are all-sky observations, the confusing sources will be
seen through all the different patches of the ionosphere on the sky,
making this a calibration problem with many degrees of freedom.  Once
more, the coming years have to show down to which frequencies
ground-based arrays can calibrate out the obstacles imposed by
ionosphere, interference, and foreground. Current experiments do not
even attempt to push into this high-redshift range, since
well-calibrated observations at frequencies $\nu<30\,$MHz
corresponding to 21-cm emission from $z\geq 50$ seem completely out of
the question for ground-based observations.

As the ultimate goal one could imagine an actual imaging experiment of
the dark ages, delivering ``all one needs to know'' about the
Universe. However, as we have seen here, the required collecting area
is several tens of square kilometers to achieve the necessary surface
brightness sensitivity with arcminute-scale resolution. This puts such
a project in a very distant future.

Apart from the fundamental cosmology science goal, other topics are
also of potential interest. Clearly, Galactic and extragalactic
surveys are needed to shed light on how the Universe looks at these
wavelengths. One expects to be particularly sensitive to fossil radio
sources in clusters, and to be able to constrain the total energy
output of quasars. As shown by the survey equation (\ref{eq:tsurvey}),
an extragalactic half-sky survey with arcminute resolution at 10\,MHz
is possible within two years with an array of a few hundred elements
and baselines up to 100\,km down to a 5 $\sigma$ detection of the
confusion background.

A fairly unique science application at low frequencies would be the 3D
tomography of the local warm interstellar medium. Since the hot gas in
our galaxies starts to become optically thick between 0.1 and 3 MHz, a
ULW radio array could see the entire spatial distribution of this gas
from the Galactic Center to the Local Bubble by simply stepping
through all frequencies. The gas will have imprinted in its structure
the recent history of the Galaxy and in particular of the solar system
environment. This is of particular interest since the Local Bubble
pressure has a direct effect on the size of the heliosphere and thus
on the cosmic ray flux on Earth and its related effects on weather and
biological mutation rates \citep{SFBea06}. Hence, one might be able to
shed some light on the co-evolution of life in the solar system and
its interstellar environment. To perform this survey, a lunar
telescope would require several tens to hundred dipoles, several tens
of kilometer baselines and a broad frequency coverage down to 0.1
MHz. That could be an realistic science goal for the first lunar
telescope installations.

Another new application in ULW astronomy is the search for transient
sources. Here very little is known \emph{a priori}. Jupiter is the
prime example with its cyclotron maser flares at 10 MHz which are
completely invisible at frequencies above 40\,MHz due to a sharp
spectral cutoff, depressing the flux by 5 orders of magnitude
\citep{Zarka07}. It has been suggested that this could be used for the
detection of the magnetospheres of exoplanets
\citep{LFDea04,GMMea05,GZS07}, but will certainly also be of interest
to investigate solar flares in nearby stars. Transient detection
requires good instantaneous $(u,v)$~coverage, tens of thousands of dipoles,
some spatial resolution (baselines of 10-100 km), triggering
mechanisms, and buffering electronics.

Due to its huge mass and lack of atmosphere, the moon is also a unique
ultra-high energy cosmic ray detector. This is currently being used by
ground-based experiments which exploit the fact that particle cascades
in the lunar regolith should produce strong radio Cherenkov
emission. At the very highest energies for cosmic rays this promises
by far the most effective detection mechanism \citep{SBBea06}. The
main limit at present are systematic uncertainties of lunar
properties.  These could be overcome with the help of a relatively
small lunar surface radio array which would make in-situ measurements
of these cosmic-ray-induced radio flashes. Significant advances might
already be possible with a first prototype array of only a few
dipoles. Of particular interest is the transparency of the lunar
regolith below 10 MHz. If this continues to increase with decreasing
frequency, already a few dipoles could provide an effective neutrino
detector volume of many cubic kilometers in the future. A by-product
of such impulsive radio noise measurements could be the measurement
and localization of meteoritic impacts on the moon.

\subsection{Open questions}

It is obvious that a full-blown lunar ULW radio array with thousands
of elements cannot be the first step.  Indeed, there are a number of
open questions about observability and design constraints for a
moon-based ULW radio telescope, which need to be answered before a
design for a lunar array can be contemplated in detail, let alone be
finalized:
\begin{enumerate}
\item What are the properties of the lunar ``ionosphere'', in
  particular at the likely sites for a lunar array on the far side
  and near the poles? 
\item What are the subsurface and electrical properties of the moon?
\item How strong is the shielding of terrestrial radio waves the in polar region?
\item How strong it the impulsive radio background noise on the surface?
\item How strong are electromagnetic effects associated with meteorite
  impacts that could cause impulsive radio noise on the lunar surface?
\item \label{en:oq.scatt} Are the IPM and ISM scattering
  extrapolations from higher frequencies still accurate near 3 MHz? 
\item \label{en:oq.ism} What is the detailed synchrotron and free-free
  emissivity and optical depth of the ISM, as a function of galactic coordinates?
\item \label{en:oq.ion} What is the lowest frequency at which
  ionospheric fluctuations above a terrestrial observatory can be
  calibrated sufficiently well to achieve the dynamic range necessary
  for imaging faint sources? 
\end{enumerate}
Some of these questions will be answered by LOFAR or MWA in the near
future, followed later by the SKA. Instruments on current-generation
lunar orbiters like the on-going Japanese KAGUYA/SELENE mission will
also provide new insights into questions about the electrical
properties of the moon. Nevertheless, some questions will remain that
will require measurements and prototyping on the moon directly.

\subsection{Stepping stones \emph{en route} to a large lunar array}

The big advantage of the phased dipole array technology is that it is
robust, very flexible, and scalable with the number of
antennas. Therefore, one can easily define a program with several
steps signified by a growing number of antennas. Dipole antennas can
in principle be packaged in a small device and therefore could initially
piggy-back on other missions for feasibility studies. The following
steps could be envisaged:

\begin{enumerate}
\item a dedicated pathfinder experiment with a single antenna to 
\begin{itemize}
  \item analyze the transmission of the lunar ionosphere
  \item investigate the surface and subsurface electrical properties
    of the Moon
  \item determine the beam shape and stability of a dipole on the surface
  \item monitor the  radio noise environment on the lunar surface
  \item measure the redshift of the global Epoch of Reionization
    (\S\ref{s:scidrv.cosmo.gEoR})
\end{itemize}
\item a two-element interferometer with a moving antenna that can
\begin{itemize}
\item demonstrate lunar interferometry
\item perform a first simple ULW all-sky survey
\end{itemize}
\item a small lunar ULW telescope consisting of $\ge3$ dipoles in the
  100 MHz regime that can
\begin{itemize}
\item act as cosmic-ray, neutrino, and meteoritic impact detector
  (\S\S\ref{s:scidrv.uhe.cosmics}, \ref{s:scidrv.meteo})
\item demonstrate lunar interferometry
\end{itemize}
\item an imaging telescope operating in the 10\,MHz regime with 30--300
  dipoles distributed over 30--100\,km, for
\begin{itemize}
\item extragalactic surveys of large-scale radio structures in
  clusters, radio galaxies, etc.\ (\S\ref{s:scidrv.extragal})
\item tomographic mapping of the Galactic ISM
  (\S\ref{s:scidrv.galaxy.ism})
\item monitoring planetary radio bursts (\S\ref{s:scidrv.transients.planets})
\item neutrino and cosmic ray measurements (\S\ref{s:scidrv.uhe})
\end{itemize}
\item a full-size array optimized for the 10--100\,MHz regime, with
  $10^3-10^7$ antennas spread over tens of kilometers to 
\begin{itemize}
\item investigate the cosmic dark ages (\S\ref{s:scidrv.cosmo})
\item search for transient Galactic sources, in particular exoplanets (\S\ref{s:scidrv.transients.planets})
\end{itemize}
\end{enumerate}

Among the early prototypes, a simple two-element interferometer near
one of the poles is a particularly appealing concept. It could consist
of a fixed antenna and one installed on a rover, with the rover
operating over several months and slowly moving further away. The
combination of lunar rotation and motion of the rover out to tens of
kilometers would provide a decent $(u,v)$~coverage and allow one to perform
an interesting imaging survey already with a small mission.

\subsection{Potential realizations}

The realization of these arrays is a matter of current study. Dipoles
are the preferred primary element for the simplicity and ease of
deployment, but also magnetic loop antennas and tripoles are possible
options. Since the Moon's surface is a much better insulator than the
Earth's, in principle the dipoles could be placed directly on the
lunar surface, without a defined ground plane \citep{Takahashi03}.

To keep data rates for earth-moon telemetry manageable (details
below), an array with a large number of elements requires a correlator
to be placed on the moon.  For the deployment of a large number of
antennas, \citet{JLMDea07} have considered printing of antennas onto
very thin foil that can be rolled out by astronauts. However, for a
smaller number of dipoles robotic deployment by rovers or even simple
ballistic ejection mechanism seem to be sufficient.

A major concern is the power supply to individual antenna
elements. This has to be achieved either locally, through small solar
cells and batteries for some fractional night operation, or through
wire connections to a central solar power generator. The limited
capacity of current rechargeable batteries is a major limitation for
any night operation. Day-time operation has always the danger of being
affected by strong interference from the sun and the lunar ionosphere
at the very longest wavelengths.

A second challenge is the data connection between the individual
elements and the central data processor (``correlator'') on the Moon,
where the data streams from all antennas have to be correlated with
each other, and the connection between the correlator and the ground
station on Earth.  For 10\,MHz bandwidth and 8-bit digitization at
20\,MS/s a connection of $ \approx 200$ Mbit/sec is needed. Rather
than digitizing the radio waves at the dipoles and risking
self-generated interference, it may be preferable to carry out the
digitization close to the correlator. For this, the analog radio
signal could be modulated onto a laser or a glass fibre
connection. Due to the curvature of the moon and its rocky nature,
laser connections will only work over a limited distance. Wireless
radio communication at high frequencies could also be considered, but
this option is disfavored as it would disturb the radio-quiet nature
of the site.  The data rate demands between the correlator and Earth
are much less than the intra-lunar connections, but have to go through
a relay satellite for a location on the far side.

The data correlator itself does not appear to be a fundamental
problem. Modern general-purpose Field Programmable Gate Arrays
(FPGA)-based correlators
\citep[e.g.,][]{vonHerzen98,WBBea04,TKDea05,PBCFea06,PBCea07} are
already able to integrate the cross-correlation of 30 antennas with 10
MHz bandwidth essentially on a single chip.

\subsection{Conclusions}
The low frequency end of the electromagnetic spectrum is currently not
well-explored, and the very-low frequency end is indeed one of the
last portions of the electromagnetic spectrum to remain terra
incognita in astrophysics. A lunar low-frequency array would therefore
enable pioneering observations in a number of scientific areas
(Tab.~\ref{t:overview}), from the Dark Ages in cosmology to exoplanets
and astroparticle physics. The field is still in its infancy and
serendipity may hold some surprises. After all, the transition to long
wavelengths also marks a qualitative transition, where coherent
processes play a much stronger role and hitherto unknown ultra-steep
spectrum source could reveal themselves for the very first time.  If
lunar exploration becomes a reality again in the next decades, then a
low-frequency array for astrophysics and space science should be an
important part of it, supplementing the impressive array of new radio
telescopes on the ground.

\section*{Acknowledgements} Early parts of this work were supported by
ASTRON and EADS Astrium within a joint concept study for a lunar
infrastructure for exploration. We are thankful to G.\ Woan,
J.\ Noordam, A.\ Gunst, S.\ Wijnholds, J.\ Bregmann, M.\ Meijers,
H.\ Butcher, M.\ Garret, D.\ Werthimer, J.\ Lazio, H.J.\ Heidmann,
H.\ Guenther, P.\ Zarka, S.\ Zaroubi, and Hartmut M\"uller, as well as
the participants of the lunar observatories workshops in
Bremen\urlfn{http://www.astron.nl/moon} for many interesting
discussions on this and related topics. We thank Rick Perley for
discussions on the survey equation and wide-field imaging, Tom
Maccarone for the pulsar idea, and Benedetta Ciardi for discussions of
the global 21-cm signal from the dark ages and providing the input
data for Figs.~\ref{f:global21cm} and \ref{f:darkages-line}.  We are
grateful to the anonymous referee for constructive comments which have
allowed us to improve this review.


\end{document}